\documentclass[twocolumn,astrosymb,trackchanges]{aastex7}

\shorttitle{RS CVn Binaries in VLASS and RACS}
\shortauthors{Rao et al.}

\begin{document}

\title{Radio Properties of RS Canum Venaticorum Variables in VLASS and RACS}

\author[0000-0002-1349-8173]{Suhasini S.~Rao}
\affiliation{University of Alberta,
CCIS 4-181, Edmonton, Alberta, Canada}
\email[show]{suhasin1@ualberta.ca}  

\author[0000-0001-6682-916X]{Gregory R.~Sivakoff}
\affiliation{University of Alberta,
CCIS 4-181, Edmonton, Alberta, Canada}
\email{sivakoff@ualberta.ca}

\author[0000-0003-3944-6109]{Craig O.~Heinke}
\affiliation{University of Alberta,
CCIS 4-181, Edmonton, Alberta, Canada}
\email{heinke@ualberta.ca}

\author[0009-0001-3587-6622]{Swetha Arumugam}
\affiliation{Department of Physics and Astronomy, Clemson University, Clemson, SC, USA}
\email{swethaa@clemson.edu}

\author[0000-0003-0699-7019]{Dougal Dobie}
\affiliation{Sydney Institute for Astronomy, School of Physics A28, University of Sydney, NSW 2006, Australia}
\email{d.dobie@sydney.edu.au}

\author[0000-0002-4405-3273]{Laura Driessen}
\affiliation{Sydney Institute for Astronomy, School of Physics A28, University of Sydney, NSW 2006, Australia}
\email{laura.driessen@sydney.edu.au}

\author[0000-0001-6295-2881]{David L.~Kaplan}
\affiliation{Center for Gravitation, Cosmology, and Astrophysics,
Department of Physics, University of Wisconsin-Milwaukee, Milwaukee, WI, USA}
\email{kaplan@uwm.edu}

\author[0000-0002-9994-1593]{Emil Lenc}
\affiliation{Australian Telescope National Facility, CSIRO Astronomy and Space Science,
Epping, NSW, Australia}
\email{emil.lenc@csiro.au}

\author[0000-0003-1575-5249]{Joshua Pritchard}
\affiliation{Sydney Institute for Astronomy, School of Physics A28, University of Sydney, NSW 2006, Australia}
\email{joshua.pritchard@sydney.edu.au}

\begin{abstract}
We performed a systematic search for radio emission from RS Canum Venaticorum (RS CVn) binaries, selected from the International Variable Star Index (VSX) catalog, in the Very Large Array Sky Survey (VLASS; three epochs) and Rapid ASKAP Continuum Survey (RACS; two epochs) data. We detected 108 candidate radio-emitting RS CVn in at least one epoch. Several of these systems rank among the most radio-luminous RS CVn binaries reported to date. The radio and X-ray luminosities, obtained from cross-matching with the eROSITA and ROSAT X-ray catalogs, are consistent with the G{\"u}del--Benz relation for magnetically active stars, but are also comparable to radio-luminous quiescent black hole X-ray binaries, indicating a potential for misidentification between these two classes. Analysis of optical, radio, and stellar properties indicates that optically bright RS CVn (i.e., those with at least one giant component) are radio-quieter and have periods that are consistent with lower coronal activity. However, two of these optically bright RS CVn systems show persistent and unusually high radio specific luminosities ($L_{R,\nu}>2\times 10^{17} {\rm \, erg \, s^{-1} \, Hz^{-1}}$) across all observed epochs, showing that stellar activity can produce relatively persistent radio signals as bright as quiescent black hole binaries.
\end{abstract}


\section{Introduction}\label{sec:introduction}
Advancements in radio and X-ray astronomy, along with the advent of wide-field surveys, have added to the ever increasing population of transient and variable sources being detected \citep[e.g.][]{Tetarenko_2016, BlackCAT_2016, Abdollahi_2017, Matthews_2019, Driessen_2024, Ayanabha_2024, Grotova_2025, Murphy_Kaplan_2026}. These transients range (in mass) from chromospherically active M-type stars to flaring active galactic nuclei (AGNs). Interestingly, most classes of stellar sources exhibit correlated radio and X-ray luminosities, likely shaped by the underlying physical properties of the system. For example, accreting compact objects, such as X-ray binaries (XRBs), have an X-ray--radio correlation that is generally attributed to accretion disks and relativistic jets \citep{GalloFenderPooley_2003, Gallo_2006}. In contrast, systems with non-degenerate stellar components with convective envelopes, such as M-type stars, BY Draconis binaries, FK Comae stars, and RS Canum Venaticorum (RS CVn) binaries, display a correlation resulting from chromospheric activities and coronal heating \citep{Drake_1989, Guedel_1993, Gudel_Mdwarf_1993, Williams_2014}.

RS CVn binaries are among the most luminous stellar radio sources that are not associated with compact objects. These chromospherically active F--K type binaries, typically of luminosity classes II--IV \citep{Hall_1976}, are known for their intense magnetic activity, enhanced coronal X-ray emission, and strong optical variability. They are characterized by their short orbital binary periods, typically ranging from $\sim$1 to 30 days, strong \ion{Ca}{2} H and K emission lines, and strong H$\alpha$ Balmer lines. The high magnetic activity of these binary systems is attributed to rapid stellar rotation from tidal locking \citep{Walter_1981}, which often gives rise to large starspots that can sometimes cover up to $\sim$40\% of the stellar surface \citep{Rodono_1995}. A large fraction of the optical photometric variability of the star is attributed to the spots rotating in and out of view \citep{Eaton_1979, Kimble_1981}. Since both individual charged particles and bulk plasma emission emit at radio frequencies, radio observations are particularly sensitive to magnetic fields and are therefore excellent tracers of chromospheric and coronal activities commonly associated with these systems \citep{Dulk_1985, Bastian_1998, Gudel_2002}. 

Radio emission from an RS CVn system is believed to originate from: (1) magnetically active regions of the stellar corona of one or both components of the binary; and/or (2) the magnetic interaction between the components \citep{Slee_2008}. For example, a VLBI observation of an RS CVn binary revealed both a core and halo structure of the radio emission \citep{Mutel_1985a}, suggesting both origins can simultaneously play a role. Regardless of the precise origin of the radio emissions, most earlier studies \citep[e.g.][]{Mutel_1985b, Mutel_1987, Morris_1988, Drake_1989} observing these binaries at GHz frequencies interpreted the observed un-polarized or weakly polarized emission with low brightness temperature as incoherent gyro-synchrotron emission. More recent studies \citep{White_1995, Slee_2008, Pritchard_2021, Vedantham_2022} have reported some of the emission has large circular polarization fractions and high brightness temperatures, suggesting coherent emission mechanisms like plasma emission or electron-cyclotron maser emission (ECME) to be responsible for the observed properties.

Previous studies have reported RS CVn systems with radio luminosities (estimated as $\nu L_{\nu}$) reaching $\sim 10^{27}$ erg s$^{-1}$ at 144~MHz and 5~GHz \citep{Drake_1989, Guedel_1993, Toet_2021, Vedantham_2022}. At radio luminosities above this, there are generally two classes of objects: high-mass (often persistent) sources like blue supergiants and Wolf-Rayet systems (isolated stars, or colliding wind binaries) \citep{Driessen_2024}; and highly variable sources, especially those with accreting compact objects \citep{Pietka_2015}. Accreting compact objects (cataclysmic variables --- CVs; and X-ray binaries --- XRBs) involve white dwarfs, neutron stars or black holes, accreting material from nearby companion stars. For XRBs and dwarf-nova CVs in outburst, the radio emission is thought to come from a relativistic jet, while the X-ray emission originates from the inner parts of the accretion disk or the jet \citep{Markoff_2005,Coppejans2020}.
If flaring RS CVn binaries can have similar radio and X-ray luminosities as CVs and quiescent XRBs, the risk of confusion between these classes significantly increases in the absence of multi-wavelength data that can confirm the nature of the objects \citep[e.g.,][]{Shishkovsky_2018}. RS CVn systems can, in principle, be differentiated from the accreting compact objects by their optical properties, with the presence of optical periodic modulation and narrow chromospheric \ion{Ca}{2} H and K lines pointing towards an RS CVn nature. However, in the absence of such data (often true in the case of globular clusters --- GCs), astronomers have identified candidate quiescent XRBs based on their radio and X-ray brightness alone \citep[e.g.,][]{Strader_2012, Chomiuk_2013,Tetarenko16b}. Accurately distinguishing active binaries like RS CVn from quiescent XRBs is essential in constraining the number of stellar-mass black holes retained in GCs \citep{Zhao_2021, Tudor_2022}, with implications for models of dynamic interactions within GCs \citep[e.g.,][]{Heggie_Giersz_2014}. 

In this work, we report the detection of some of the most radio-luminous RS CVn binaries observed to date, further complicating the distinction of chromospherically active binaries and accreting compact object systems based solely on radio and X-ray properties. We also explore which stellar properties affect the radio luminosity and/or radio detectability of these RS CVn binaries. The paper is structured as follows. In Section \ref{sec:sampleSelection}, we discuss the selection of a sample of candidate RS CVn systems from VSX and \textit{Gaia}. In Sections \ref{sec:Radio_and_Contamination} and \ref{sec:XrayObs} we discuss the source finding methods, contamination rejection,  reliability in radio surveys, and cross-matching to X-ray surveys. In Sections \ref{sec:results} and \ref{sec:discussion}, we discuss our results and their implications.

\section{Sample Selection} \label{sec:sampleSelection}
\subsection{Catalog Selection and Gaia cross-match} \label{subsec:cat&gaia}
To accurately calculate luminosities (using optically-derived distances) and improve our astrometric identifications, we first matched a large sample of RS CVn systems with \textit{Gaia}. We used the International Variable Star Index (VSX) catalog \citep{Watson_2006} from the American Association of Variable Star Observers (AAVSO) to select all sources flagged as RS CVn variables (\texttt{Type} = \texttt{RS}). The VSX catalog is a collection of variable stars collected from user submissions, various catalogs, and literature. Many VSX identifications are first suggested by citizen astronomers, and then approved by VSX Data Moderators or the VSX Project Administrator \footnote{\url{https://vsx.aavso.org/index.php?view=about.top}}. The sources flagged as RS CVn type are primarily identified based on photometric variability and therefore do not necessarily satisfy all criteria for the classical definition of an RS CVn system. Since we cannot confirm that every VSX-identified RS CVn system meets the full definition of an RS CVn system, we treat all VSX-identified RS CVn systems as RS CVn candidates. We used the 2022 peer-reviewed version of the VSX catalog (Version 2022-05-30) on Vizier, which includes 73,332 RS CVn candidates, as our base sample.

To identify \textit{Gaia} counterparts for the VSX sources, we cross-matched our base sample with the \textit{Gaia} Data Release 3 (DR3) catalog \citep{Gaia_2023}. We initially chose a generous matching radius of $10\arcsec$ to account for the proper motion of the stars. Given that the VSX catalog does not list the observation epochs associated with its J2000-equinox coordinates, we treated all VSX source positions as J2000-epoch positions. With this, we found 73,329 sources with at least one \textit{Gaia} match; we excluded VSX RS CVn candidates lacking a \textit{Gaia} counterpart from further analysis. For all the sources in the matched sample, we propagate (with proper motion corrections) the matched Gaia positions from their J2016-epoch to J2000-epoch and selected the nearest \textit{Gaia} source as the VSX counterpart.

To account for high proper-motion systems that might fall outside the initial 10$\arcsec$ search constraints, we conducted a supplemental match of all epoch-corrected \textit{Gaia} DR3 sources with proper motion $>500 \rm{\, mas \, yr^{-1}}$ with the whole VSX RS CVn catalog using a matching radius of $1\arcsec$ to find better/additional matches. This method recovered one additional RS CVn candidate and identified a more accurate (closer) \textit{Gaia} counterpart for a previously matched source. After replacing the latter with the improved match, our final sample consisted of 73,330 sources. Two sources without a \textit{Gaia} match --- $\alpha$ Aur and $\gamma$ Leo --- are too optically bright to be detected by \textit{Gaia}, and are therefore removed from further consideration in this study\footnote{Neither source was detected in our radio data.}.

We extracted the \textit{Gaia} astrometric and photometric measurements for this sample. Throughout our work we used the standard inverse parallax method to measure the distances. Comparison with the probabilistic distance estimates from \citet{BailerJones_2021} shows that the two estimates differ by $\le6\%$ for $90\%$ of our sample, as expected for nearby ($<$2 kpc; see below) sources. We removed 321 sources that were missing a $G_{BP}-G_{RP}$ color or $G$-band magnitude in \textit{Gaia}. To ensure reliable distance measurements and be consistent with \cite{Leiner_2022}, we rejected 55,196 more sources whose \textit{Gaia} parallax signal-to-noise (SNR) was less than $5$, or that had a distance beyond $2 {\, \rm kpc}$.

This yielded a final VSX-\textit{Gaia} sample of 17,813 RS CVn candidates with unique \textit{Gaia} counterparts.

\begin{figure}[t]
    \plotone{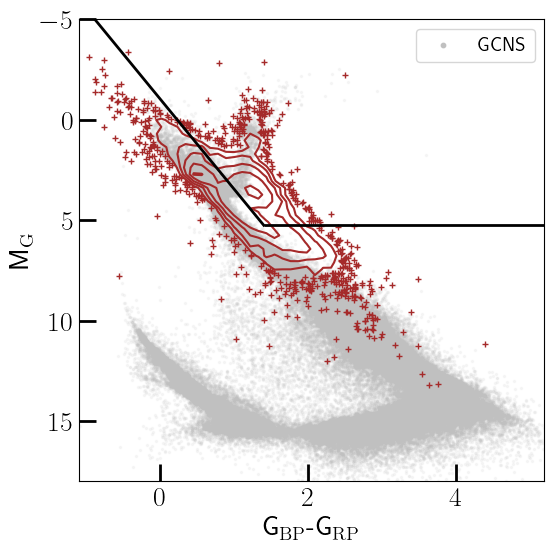}
    \caption{Extinction-corrected CMD of the VSX-\textit{Gaia} sample of RS CVn candidates. The sample is represented as contours in regions of source density $>8$ with contour levels increasing by factors of 2. Below this source density, sources are individually marked with crosses. The sample has a median E($G_{BP}-G_{RP}$) of $\sim$0.4 and $A_G$ of $\sim$0.8. The color-magnitude cuts to select the final sample of RS CVn candidates are shown as black dashed lines. The final sample selected lie in the redder and brighter side of the cuts. This catalog has a median E($G_{BP}-G_{RP}$) of $\sim$0.3 and $A_G$ of $\sim$0.7. The \textit{Gaia} Catalog of Nearby Stars (GCNS) sources are shown in gray for reference.
    \label{fig:selection}}
\end{figure}

\subsection{Extinction Correction and Color-Magnitude Cut} \label{subsec:extinctionCMcut}

RS CVn systems can overlap with the positions of T Tauri stars and high-mass main-sequence stars in the color-magnitude diagram (CMD), especially when the stars are heavily obscured by dust. To correct for this, we applied extinction corrections using the Bayestar 3D dust map \citep{Green_2019} for sources above a declination ($\delta$) of $-30^\circ$ and the DECaPS 3D dust maps \citep{Zucker_2025} for those below. We then derived the color excess values ($E(g-r)$ from the Bayestar map and $E(B-V)$ from the DECaPS map) at each source position and \textit{Gaia}-based distance. We converted the color excess to $A_V$ assuming $R_V = 3.1$ \citep{SF_2011}. For 123 southern sources without valid estimates, we adopted the $A_V$ of the nearest source in our VSX-\textit{Gaia} sample. Given a median separation of $\sim 110$ pc and the high Galactic latitudes ($b \geq 10^\circ$) for most of these sources, this interpolation provides a reasonable estimate of the local interstellar extinction. Using the precomputed \texttt{dustapprox} models \citep{Fouesneau_dustapprox_2022}, we iteratively calculated the extinction coefficient for the \textit{Gaia} $G$, $G_{BP}$, and $G_{RP}$ bands and corrected for the reddening until the correction within each iterative step is $<\Delta(G_{BP}-G_{RP}) \leq 0.001$. This sample has a median E($G_{BP}-G_{RP}$) of $\sim$0.4 and $A_G$ of $\sim$0.8.

The CMD of extinction-corrected RS CVn candidates is shown in Figure \ref{fig:selection}. Regions of high source density ($>8$ sources) are shown as contours with contour levels increasing by factors of 2. Sources in lower density regions ($<8$ sources) are individually marked as crosses. After applying an extinction correction, we see sources that could potentially be main-sequence, T Tauri stars, and other pre-main-sequence stars (PMS). The majority of T Tauri stars and PMS are faint and lie just above the main sequence \citep[e.g., see Figure 6 \textit{lower-left} from][]{McBride_2021}. To obtain a cleaner RS CVn sample, we applied color–magnitude cuts (shown as dashed lines), following \citet{Leiner_2022}. The diagonal cut, offset $\sim0.5$ redward from the main sequence, removes potential main-sequence contaminants. The absolute magnitude cut rejects all sources with $M_G >5.25$ mag and therefore excludes the majority of potential T Tauri stars and PMS. The final cleaned sample consists of a total of 7,252 RS CVn candidates that are redder and brighter than the color-magnitude cut-off lines. This final RS CVn candidates catalog has a median E($G_{BP}-G_{RP}$) of $\sim$0.3 and $A_G$ of $\sim$0.7. 

Since RS CVn systems are inherently binary in nature and \textit{Gaia}’s single-star astrometric fitting is known to degrade or fail for unresolved binaries \citep[e.g.,][]{Penoyre_2020, CastroGinard_2024}, we examine whether our sample selection introduces a systematic bias against binary systems. For this we use the Renormalized Unit Weight Error (RUWE), which serves as a proxy for unmodeled binarity, where RUWE $<1.4$ indicates a reliable single-star solution and RUWE $>1.4$ is likely to be a higher-order system \citep{Lindegren_2018}. To quantify selection effects, we first analyzed the 285 sources excluded by our parallax SNR $<5$ threshold. Of these, 184 sources ($\sim 64\%$) exhibit RUWE $<1.4$, indicating that their low SNR stems from poorly constrained single-star astrometry rather than unmodeled binarity. The remaining 101 sources with higher RUWE account for a negligible fraction ($\sim 1.5\%$) of our total VSX-selected sample, none of which yield radio detections (as discussed in Section \ref{subsec:radioObs}). 

We also rejected 251 sources lacking parallax, proper motion, and RUWE estimations. Assuming the worst-case scenario where all of them pass our initial RS CVn criteria, only 207 pass our VSX-\textit{Gaia} separation threshold of $< 1\arcsec$ (see Section \ref{subsec:gaia_cc} below). Moreover, enforcing our sample selection on the availability of photometric data requirement reduces these 207 sources to just 82 sources. Among these, 27 objects ($\sim 33\%$) would be fainter than our absolute magnitude completeness threshold for RS CVn system ($M_G > 5.25$ mag) at our 2-kpc distance limit, disqualifying them regardless of astrometric quality. The remaining subset (55) constitutes $<1\%$ of our final VSX-\textit{Gaia} sample, of which only one is detected in the radio (present in VLASS but undetected in RACS; discussed in Section \ref{subsec:radioObs}). Therefore, we believe that the exclusion of  sources without precise parallaxes does not introduce any significant systematic bias into our (optical and/or radio) samples.

\subsection{Gaia Chance Alignment} \label{subsec:gaia_cc}

A positional match between the VSX and \textit{Gaia} coordinates does not guarantee physical association since chance alignments are heavily influenced by the sky densities of both surveys. Since we do not know the true source densities, we adopt an empirical approach followed by \citet{Lepine_2007}, which we briefly discuss here. This approach relies on the principle that the probability of chance alignment increases with increasing angular separation between the target and the matched source. To determine the chance alignment distribution as a function of angular separation, we offset the VSX RS CVn positions in five 75$\arcsec$ steps along five different position angles and repeated the \textit{Gaia} cross-match as described in the previous two sections. The multiple offsets ensured a statistically significant sample set, while the 75$\arcsec$ displacement minimizes the risk of re-matching the same true counterpart. 

\begin{figure}[t]
    \plotone{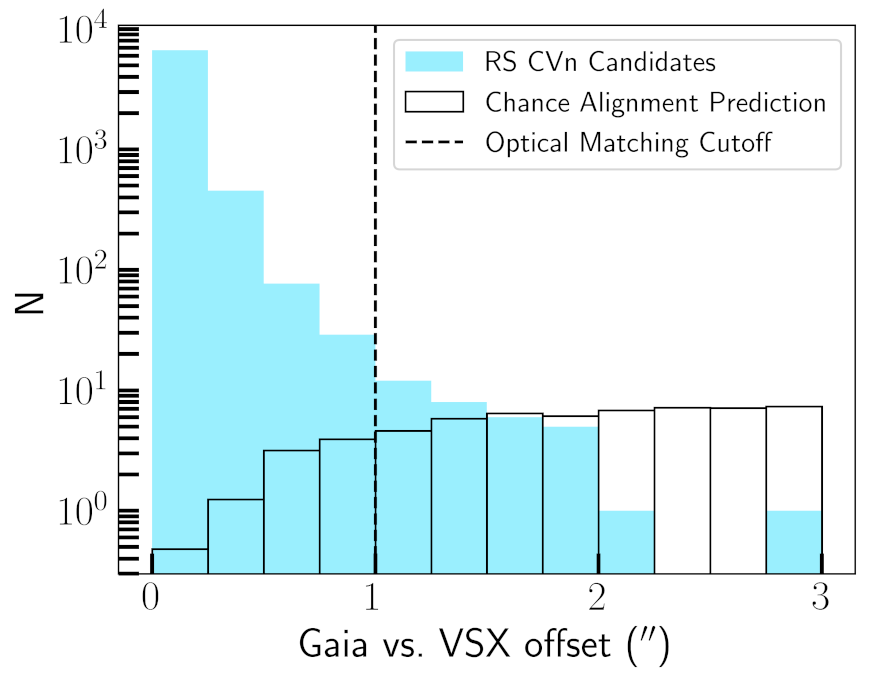}
    \caption{Histogram of the angular separation between the VSX and \textit{Gaia} coordinates. The final candidate RS CVn sample is shown in cyan. The black-line histogram shows the number of chance alignments estimated by matching 'false' coordinates with \textit{Gaia} DR3. The distributions suggest that sources with separation $> 1\arcsec$ are likely to be false matches. We remove 34 sources with angular separation $>1\arcsec$ from our sample.
    \label{fig:gaia_cc}}
\end{figure}

The distribution of these `false' matches as a function of angular separation is shown in Figure \ref{fig:gaia_cc}, where false matches dominate beyond 1.5$\arcsec$. We adopted a \textit{Gaia}-VSX separation threshold of $<1\arcsec$. The corresponding reliability for this sample, estimated using the above empirical method, is $\sim99.9\%$, and therefore nearly all are likely to be true associations. Since the stellar density as traced by \textit{Gaia} is known to drastically increase near the Galactic equator, we also evaluated the \textit{Gaia}-VSX angular separation distributions as a function of Galactic latitude. Even at low Galactic latitudes ($b$; $2^{\circ}<|b|<10^{\circ}$), where the probability of chance alignment is maximized due to high source density and relatively lower dust extinction compared to $|b|<2^{\circ}$, the probability of matching (with a matching radius of $1\arcsec$) a VSX-reported RS CVn candidate to an unrelated \textit{Gaia} source within $2\text{ kpc}$ passing our astrometric and photometric quality cuts (as defined in Section \ref{subsec:cat&gaia}) is  $\sim0.2\%$. This corresponds to an extremely high reliability of $\sim99.8\%$ for the sample with the \textit{Gaia}-VSX separation threshold of $<1\arcsec$. This is unsurprising since the vast majority of the \textit{Gaia}-VSX sample have an angular separation of $<0.5\arcsec$, well clear of the false-match distribution tail.

\begin{figure*}[t]
    \plotone{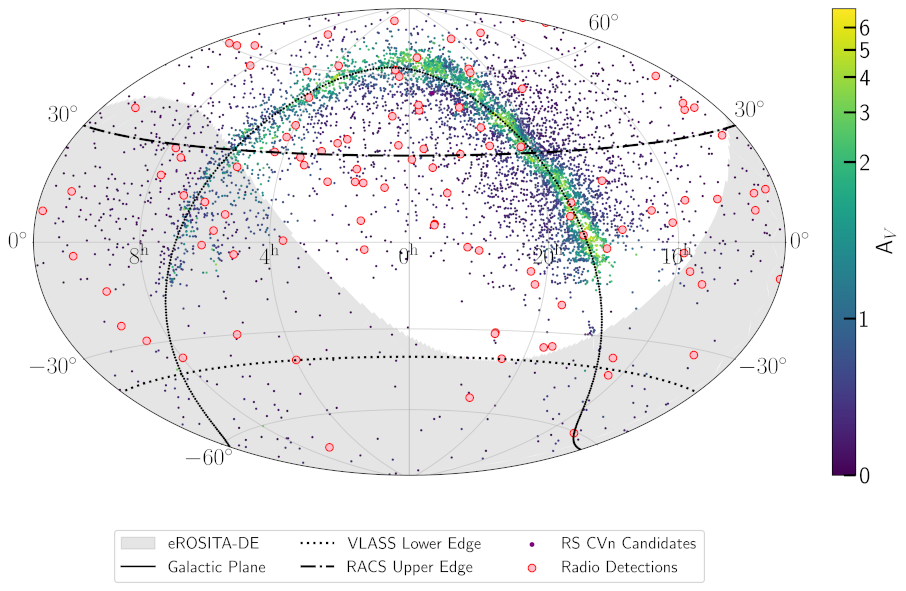}
    \caption{A projection map of the final sample of RS CVn candidates color-coded by line-of-sight extinction, $A_V$. The dotted line marks $\delta > -40^\circ$, above which VLASS provides full coverage. Similarly, the dot–dashed line marks the boundary of RACS data with good resolution at $\delta < -30^\circ$. The radio-detected sample after unrelated source contamination removal is shown in red. The gray region marks the western Galactic sky covered by eROSITA-DE. We do not explicitly show the sky coverage of the ROSAT survey since it is an all-sky survey. The bias towards the northern hemisphere and the Galactic plane is likely due to observational emphasis in those regions.
    \label{fig:skydistribution}}
\end{figure*}

Of the 7,252 sources in the parent RS CVn catalog, we rejected 34 candidates that had a \textit{Gaia}-VSX separation $>1\arcsec$, and retained 7,218 candidates. A table containing \textit{Gaia} DR3 sky coordinates, astrometric and optical properties, VSX-reported period, and the radius of the primary (as discussed in Section \ref{subsec:brightnessTemp}) of this final parent RS CVn catalog is given in its entirety in a machine-readable format (with Table \ref{tab:allRSCVn_MR} providing a description of the table's form). 
A sky distribution map of this final sample is shown in Figure \ref{fig:skydistribution}, color-coded by $A_V$. The sample has a median $A_V$ of $\sim$0.9. The overall sky distribution of the sample is biased towards the northern hemisphere and near the Galactic plane, likely reflecting the higher concentration of observers and terrestrial monitoring facilities in the northern hemisphere.

\section{Radio Data And Source Contamination} \label{sec:Radio_and_Contamination}
\subsection{Radio Data} \label{subsec:radioObs}
We searched for radio counterparts of these RS CVn candidates using data from all-sky surveys conducted by the Jansky Very Large Array (VLA) and the Australian Square Kilometer Array Pathfinder (ASKAP). Together, these surveys provided coverage of the entire sky.

We used the VLA Sky Survey (VLASS) conducted using the upgraded VLA, known as the Karl G.~Jansky Very Large Array \citep{Lacy_2020}, located in New Mexico, USA, to cover the northern sky. VLASS is a three-epoch\footnote{Half of VLASS is being observed for a fourth time, partially to replace issues with the first epoch.} all-sky survey observed in S-band (2--4 GHz, centered at 3~GHz) with 2\farcs5 angular resolution. Each epoch observed the entire sky above a $\delta$ of $-40^\circ$, covering 33,885 deg$^2$ in area, corresponding to 82\% of the entire sky. We used the VLASS Quick-Look (QL) Stokes~I sky images, designed for rapid radio transient identification \citep{Lacy_2020}. These images are divided into regions of size $1^\circ \times 1^\circ$ called \textit{subtiles} with a typical root-mean-square (RMS) sensitivity of 120 $\mu$Jy/bm. We use data from the three original epochs in our search as those observations were complete at the time of this study. A total of 7,098 candidate RS CVn in our sample are within the VLASS footprint. 

To cover the southern sky, we used the Rapid ASKAP Continuum Survey (RACS) conducted by ASKAP, located in Western Australia. We used the two epochs of the RACS-low Stokes~I sky maps observed in 744--1032 MHz range, centered at 887.5 MHz, as our primary RACS dataset, with a sky coverage of $\sim$36,200 deg$^2$ in area \citep{McConnell_2020}. RACS-low epochs 1 and 2 observe the sky below $\delta$ of $+40^\circ$ and $+50^\circ$ respectively. Each image, called a \textit{tile}, has an approximate size of $6^\circ \times 6^\circ$. Since both epochs of RACS-low were completed at the time of this study, we used data from both epochs. Since VLASS provides at least three radio observations within the same band (2--4 GHz), we chose to include only RACS-low, which provides two observations in the same band (744--1032 MHz). This excludes single epoch RACS-mid and RACS-high surveys \citep{Duchesne_2023, Duchesne_2025}. We require images of the rms that match the flux-density-image's pixel scale. RACS low-1 \citep{Hale_2021} provides low-resolution rms files that we interpolated to the higher resolution pixel size. RACS low-2 (Duchesne et al., in preparation) provides rms images matched to the  flux-density-image's pixel scale. We note that the RACS low-2 rms images were generated using a large sliding window to calculate the rms. This window size is about the same angular size as the pixel sizes in the RACS low-1 rms image. As such, we do not feel there is a large differential effect from using the two differently generated rms images. We searched for radio emission from the 3,211 RS CVn candidates in our sample that are located at $\delta<+30^\circ$, to avoid the degraded angular resolution caused by the highly elongated synthesized beam at more northern declinations.

We used the forced aperture photometry (FAP) technique, where the telescope response function is convolved with the image to estimate intrinsic flux densities. This approach ensured consistent flux density measurements across both surveys and avoided biases introduced by differences in source-finding algorithms. We used the proper motion corrected \textit{Gaia} positions for FAP and detected a total of 164 RS CVn at $\ge 5\sigma$, with 113 and 73 in at least one VLASS and RACS-low epoch respectively. 16 of these were detected in both VLASS and RACS.

\subsection{Contamination Rejection} \label{subsec:radio_cc}
\begin{figure}[t]
    \plotone{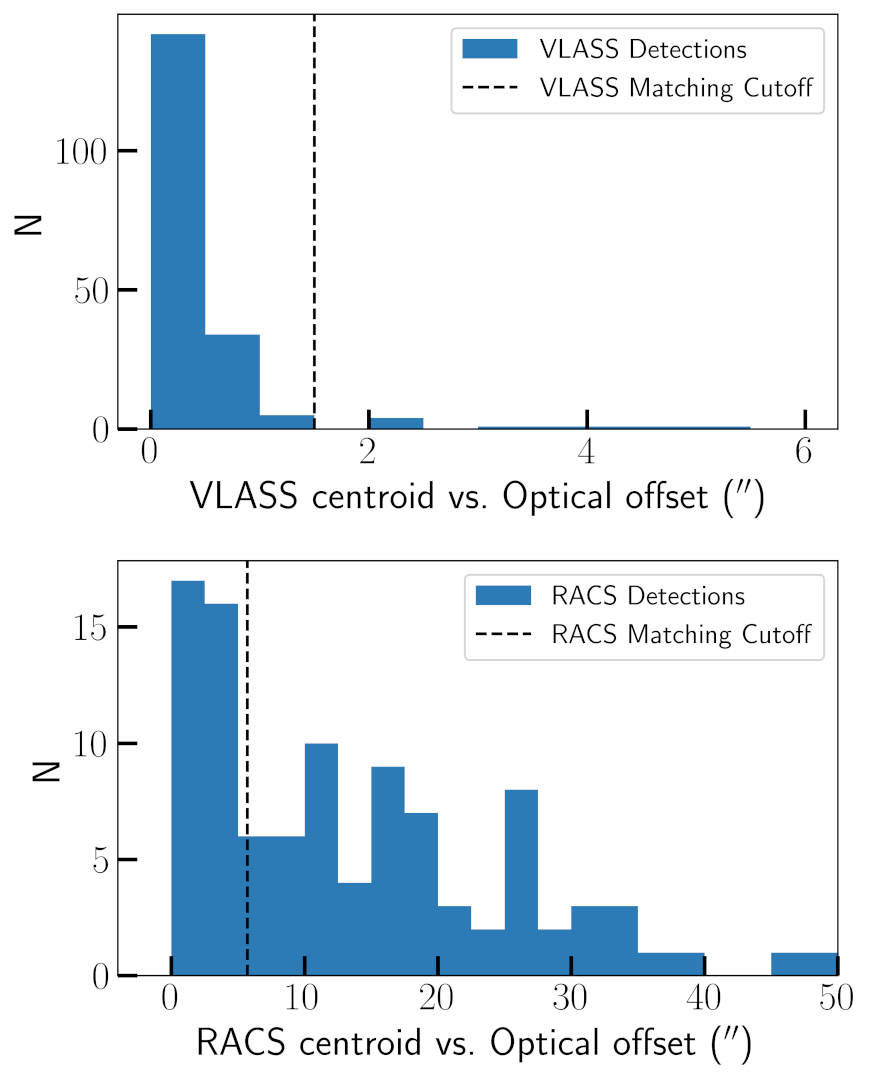}
    \caption{Histogram of the angular separation between the VSX coordinates and the estimated centroid coordinates from all VLASS (\textit{upper}) and RACS (\textit{lower}) epoch images. The astrometric matching radius threshold, shown in black dashed lines, is the 3$\sigma$ positional uncertainty of a typical $5\sigma$ source. For VLASS it is 1.5$\arcsec$,  and for RACS it is 5.7$\arcsec$. Sources with higher VSX–centroid separation were removed from the radio-detected sample (i.e., considered as non-detections).
    \label{fig:radio_cc}}
\end{figure}

Chance alignment can also occur due to the presence of background sources in the radio images. To estimate the number of false-positive detections, we first determined the centroid of all the detected RS CVn candidates in the VLASS and RACS images and calculated the separation between these centroid positions and their \textit{Gaia} coordinates. Using the median beam size of 2.85$\arcsec$ for VLASS and 17.25$\arcsec$ for RACS, and a typical 5$\sigma$ source, we estimated 1$\sigma$ statistical astrometric uncertainties of 0.3$\arcsec$ and 1.7$\arcsec$, respectively. However, for VLASS we adopted the higher uncertainty of 0.5$\arcsec$ reported by \citet{Gordon_2021a}. For RACS, we added the reported systematic uncertainty of 0.8$\arcsec$ \citep{McConnell_2020, Hale_2021} in quadrature to obtain an uncertainty of 1.9$\arcsec$. With this, we applied a 3$\sigma$ matching radius --- 1.5$\arcsec$ for VLASS and 5.7$\arcsec$ for RACS --- to distinguish associations more likely to be true from potential unrelated contaminants that happen to overlap with the forced aperture, as shown in Figure \ref{fig:radio_cc}. Of the 113 and 73 RS CVn candidates detected in VLASS and RACS (with VSX–\textit{Gaia} offset $\le1\arcsec$), 5 and 47 candidates respectively were beyond the matching radius and excluded as false radio matches (considered as non-detections). This left a sample size of 108 and 26 RS CVn candidates in VLASS and RACS respectively that were within the matching radii in all observed epochs.

\subsection{Sample Reliability} \label{subsec:radio_reliability}
To calculate the reliability of the sample, we estimated the number of chance alignments from background radio sources using a similar Monte-Carlo approach as described in Section \ref{subsec:gaia_cc}. For each VLASS and RACS epoch, we offset the \textit{Gaia} positions of the 5$\sigma$ detected radio sources by 15$\arcsec$ and 125$\arcsec$ for 150 and 50 times at 5 different position angles for VLASS and RACS respectively, and perform FAP as discussed in Section \ref{subsec:radioObs}.
We then count the number of simulated (`false') \textit{Gaia} RS CVn with $\ge 5\sigma$ radio detections in any of the 5 epochs that have their radio centroid within the appropriate matching radius (Section \ref{subsec:radio_cc}). From this we estimate a chance alignment probability of $(3.4^{+1.0}_{-0.8})\times 10^{-4}$ for VLASS and $(2.6^{+0.4}_{-0.3})\times 10^{-3}$ for RACS per RS CVn candidate within their respective footprints, where the errors indicate Gehrels uncertainties (assuming a binomial distribution). We note that the above chance alignment probabilities for our sample are $\approx 3$ and $\approx 1.6$ times higher than the chance alignment probabilities per VLASS and RACS epoch, respectively.
Given 7,098 and 3,210 RS CVn candidates within the VLASS and RACS footprints, respectively, we expect $\sim2.4$ and $\sim8.3$ chance alignments among the (108) VLASS and (26) RACS detected samples within the angular separation threshold, respectively. We thus estimate reliabilities of $\sim98\%$ and $\sim68\%$ for the VLASS and RACS radio-detected sample respectively.
The low reliability of RACS is due to the larger synthesized beam of this survey compared to VLASS, which increases the number of unrelated sources that can overlap with an aperture while performing FAP.

\subsection{Visual Inspection} \label{subsec:visualInspec}
We performed a visual inspection on the 108 and 26 RS CVn candidates in VLASS and RACS that have their radio centroid position within the matching distance of the \textit{Gaia} position, as discussed in Section \ref{subsec:radio_cc}. This resulted in the rejection of 4 sources in VLASS that were affected by imaging sidelobes and 7 RACS sources that were contaminated by coincidental background sources (mostly AGN that can be seen in VLASS) or noisy images. These numbers correspond to 2.1--7.2 and 4.4--10.8 expected `false' sources in VLASS and RACS, respectively \citep[][1-$\sigma$ Gehrels confidence intervals assuming a Poisson distribution]{Gehrels_1986}, which are consistent with the expected number of chance alignments in VLASS and RACS discussed in Section \ref{subsec:radio_reliability}. 
Removing false positive detections identified by a large radio centroid offset from the \textit{Gaia} position (5 in VLASS, 47 in RACS; see Section \ref{subsec:radio_cc}) and via visual inspection (4 in VLASS, 7 in RACS) --- amounting to a total of 9 and 54 rejections respectively --- yielded a final sample of 104 and 19 candidate RS CVn candidates in VLASS and RACS respectively. Among the 104 and 19 sources in VLASS and RACS, 14 sources were detected in both surveys. Although the rejection of the visually identified sources increases the reliability for both radio-survey samples, we retain the lower reliability estimate in Section \ref{subsec:radio_reliability} to ensure a more conservative assessment.

\section{X-ray cross-match} \label{sec:XrayObs}
We also searched for X-ray counterparts of the RS CVn candidates to complement our radio-detected sample, using the data from eROSITA instrument aboard the Spectrum Roentgen Gamma (SRG) observatory, and from ROSAT (Roentgen Satellite). eROSITA conducted a deep 0.2--10.0 keV all-sky survey known as the eROSITA all-sky survey (eRASS) \citep{Predehl_2021}. We used the first public data release, eRASS1, from the German eROSITA Consortium (eROSITA-DE), which covers the western Galactic hemisphere from December 2019 to June 2020 \citep{Merloni_2024}. The sky coverage is marked in gray in Figure \ref{fig:skydistribution}. From ROSAT, we used the revised second public released catalogue of point-like sources (2RXS) from 1990-1991 ROSAT all-sky survey (RASS) \citep{Boller_2016}. Although the radio and X-ray observations are non-simultaneous, the large sample size and the episodic nature of coronal activity make the radio and X-ray luminosity comparison valuable for understanding their flaring and non-flaring properties.

We cross-matched the proper motion corrected \textit{Gaia} coordinates of the radio-detected RS CVn candidates with the eRASS1 catalog using a 10$\arcsec$ matching radius. The matching radius is based on visual inspection of astrometric matches between eRASS1 and \textit{Gaia}/unWISE AGN catalog \citep{Merloni_2024}, and for consistency with other studies \citep[e.g.,][]{Driessen_2024}. Similarly, we use a 40$\arcsec$ matching radius with the 2RXS catalog, which is set based on the visual inspection of astrometric matches between 2RXS and Tycho 2 catalog \citep{Boller_2016}. With this, we identified 33 and 95 X-ray counterparts among the radio-detected RS CVn candidates in eRASS1 and 2RXS respectively.
All radio-detected RS CVn candidates in the western Galactic hemisphere have a match in the eRASS1 catalog. Only one radio-detected RS CVn candidate in the western Galactic hemisphere is detected in eRASS1, but not in 2RXS. The remaining 63 2RXS-matched sources are in the eastern Galactic hemisphere, where eRASS1 data are not publicly available.
We obtained the X-ray fluxes in the soft (0.2--2.3 keV) and hard (2.3--5.0 keV) bands from eRASS1, and the count rate in 0.1 –– 2.4 keV band for the 2RXS. Using CXC PIMMS\footnote{\url{https://cxc.harvard.edu/toolkit/pimms.jsp}}, we converted these to unabsorbed fluxes in 0.5––2.4 keV and 2--10 keV bands, assuming an \texttt{APEC} plasma model with a temperature of 1.2 keV, typical of RS CVn coronae, and $N_H$ derived from $A_V$ \citep[$N_H \approx 2.81 \times 10^{21} A_V {\rm \, cm}^{-2}$;][]{Bahramian_2015}.

\section{Results} \label{sec:results}
Between the VLASS and RACS-low radio surveys, we detected a total of 108 RS CVn candidates within 2 kpc out of 7218 VSX-\textit{Gaia}-matched RS CVn candidates. The radio-detected sources are marked in red in Figure \ref{fig:skydistribution}. Of the 108 sources, 96 radio detected RS CVn candidates have at least one X-ray counterpart from the 2RXS and/or eRASS1 catalogs. Twelve of the radio-detected RS CVn binaries have no detected X-ray counterpart; these are in the eastern Galactic hemisphere, where ROSAT data is not as deep as eROSITA, which is not publicly available. These radio non-detections could also be due the inherently transient nature of many RS CVn, coupled with the non-simultaneous nature of the radio and X-ray observations.

The radio and X-ray properties, along with centroid radio positions from the radio epoch with the brightest detection, of the 108 radio detected RS CVn is given in a machine readable formatted table in the online article; we describe its columns in Table \ref{tab:validRSCVn_MR}. For these sources, we collate the optical, radio, and X-ray properties, the VSX-reported period, the \textit{Gaia}-reported radius (see Section \ref{subsec:brightnessTemp}), and radio variability indices (both cross-survey and survey-specific; see Section \ref{subsec:measuredVar}) in Table \ref{tab:validRSCVn_radioXray} (these values are used throughout the figures of this paper). In Table \ref{tab:validRSCVn_radioXray}, we only list the specific radio luminosities and the soft band X-ray luminosities from the brightest detections, respectively. We use the ``RSH'' column of the parent catalog, as described in Table \ref{tab:allRSCVn_MR}, to provide matching indices to Table \ref{tab:validRSCVn_MR}.

\subsection{Measured Radio Variability} \label{subsec:measuredVar}
\begin{deluxetable}{lccc}
\tabletypesize{\scriptsize}
\tablecaption{Distribution of radio-detected RS CVn candidates based on their detection frequency across both radio surveys, and based on their variability index}
\label{tab:variabilityStats}
\tablehead{
\colhead{ } & \colhead{Single-Epoch} & \colhead{Multi-Epoch} & \colhead{All-Epoch}
} 
\startdata
High Variability & 2 & 3 & 1 \\
Moderate Variability & 13 & 27 & 14 \\
Low Variability & 0 & 0 & 1 \\
Undetermined & 37 & 9 & 1 \\
\hline
Total & 52 & 39 & 17 \\
\hline
Median SNR [$1\sigma$ CI] & 8.1 [5.5-19] & 9.8 [5.9-25] & 24 [6.7-75] \\
\enddata
\tablecomments{``Multi-Epoch'' refers to sources detected in more than one but not all available VLASS and/or RACS epochs, as compared to ``All-Epoch'' which are sources detected in all available epochs across both surveys.}
\tablecomments{``High Variability'' class have $V_L \geq 10$; ``Moderate Variability'' class have $2 \leq V_L < 10$; ``Undetermined'' class have $V_L < 2$; ``Low Variability'' class have $V_L < 2$ but  have detection $\text {SNR}>80$ in all detected epochs.}
\end{deluxetable}

The 108 radio-detected RS CVn candidates showed varied levels of detection frequencies across both VLASS and RACS: 52 sources were detected in only one epoch (``single-epoch''), 39 were detected intermittently (more than once but not in all available epochs; hereafter ``multi-epoch''), and 17 were detected across all available epochs (``all-epoch''). We quantify their variability across both radio surveys using a variability index, $V_L = L_{\rm max} / L_{\rm min}$, by assuming a flat spectral index between RACS-low and VLASS.
The most luminous radio detections (i.e, the highest specific radio luminosity) are  recorded as $L_{\rm max}$. When determining the minimum detected or detectable luminosity, ($L_{\rm min}$), we consider the measured specific luminosities when the sources are detected in a given available epoch, and the $5\sigma$ upper limits otherwise for a given epoch. Since one of the main radio emission mechanisms in RS CVn binaries is gyrosynchrotron emission, which is typically flat at the VLASS and RACS observing frequencies, a cross-survey comparison of luminosities in reasonable. Using a flat spectral index also has the advantage that comparing specific luminosities is equivalent to comparing luminosities. For completeness, we also calculate survey-specific variability indices and list them in Table~\ref{tab:validRSCVn_radioXray}. However, we restrict the discussion to the cross-survey variability index.

Sources with $V_L \geq 10$ are categorized as `high variability', while those with $2 \leq V_L < 10$ are designated as `moderate variability'. Sources with $V_L < 2$ were originally classified as `undetermined'; however, visual inspection revealed one source, BH CVn, that was detected with $V_L = 1.9$ and had $\text {SNR}>80$ in all three VLASS epochs (it was not in the RACS footprint). We have placed this source it its own category of `low variability'. The distribution of the radio-detected RS CVn candidates in these four variability classes is summarized in Table \ref{tab:variabilityStats}, along with the median signal-to-noise ratio (SNR) and the $68\%$ confidence interval ($1\sigma$ CI) of the different detection-frequency classes. 

Most RS CVn candidates showed evidence of radio flux variability of at least a factor of two (60/108). At this level of variability, and the associated SNRs, the variability in these sources is exceptionally unlikely to arise from noise in the images.
Typically, high amplitude radio flux variability in RS CVn systems is attributed to intense stellar magnetic activity, especially in the form of short-timescale (seconds to days) transient flare-like events.
However, with at most five epochs over seven years, we do not have the temporal coverage to clearly indicate a short-timescale flare. At one extreme, every radio detection could be from serendipitous observations of a stellar flare. On the other extreme, all variability that we observe could be due to persistent emission, some of which can be highly variable. Given our limited data, we only attempt interpretations in a few extreme cases, those with the greatest variability.

\begin{figure}[t]
    \plotone{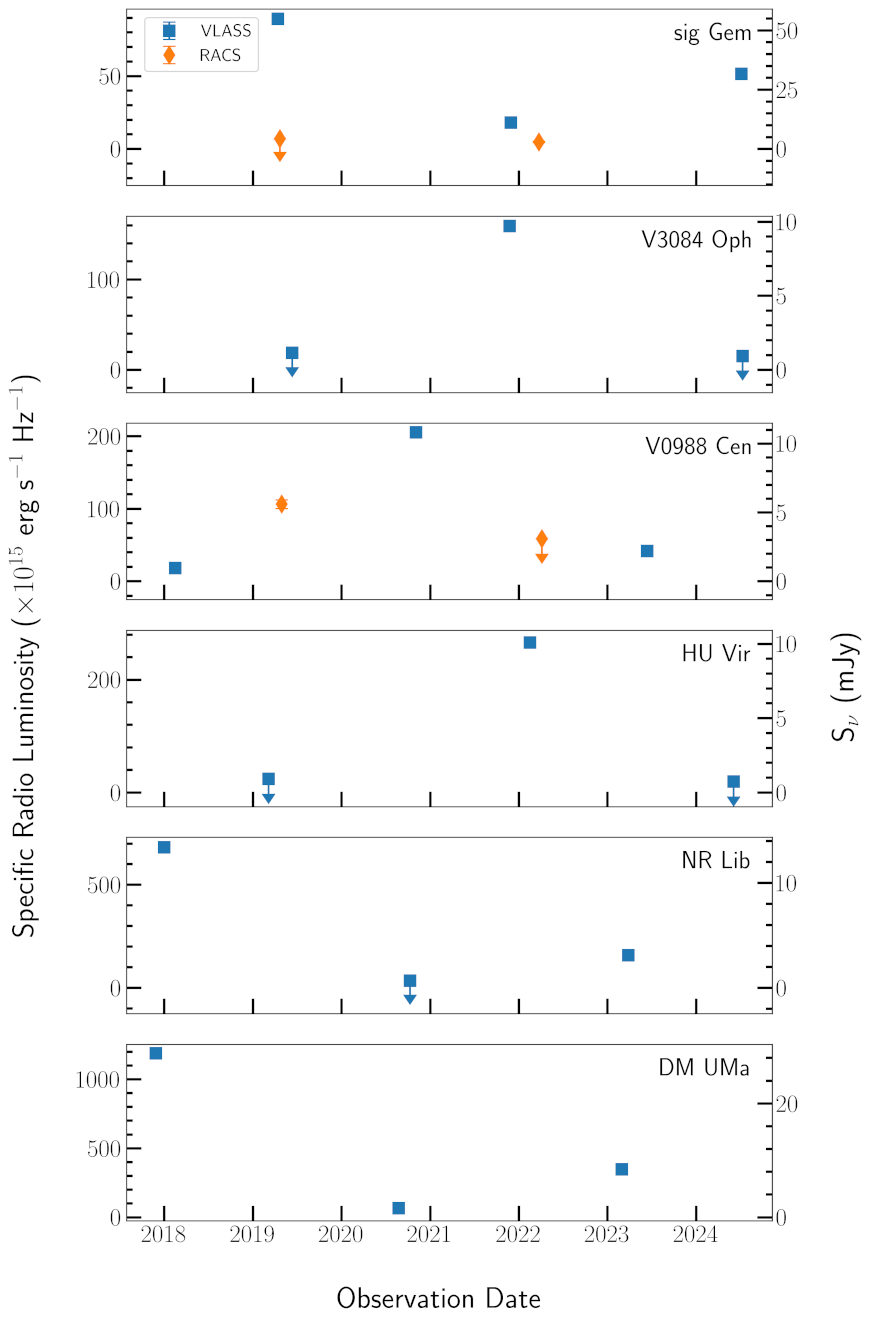}
    \caption{VLASS/RACS-low lightcurve of the six ``clearly flaring'' radio-detected RS CVn candidates (the errors for the detected sources are too small to be displayed). Upper limits ($5\sigma$) are shown for each of the non-detections in the VLASS and/or RACS-low epochs. VLASS observations are shown in blue and RACS observations are shown in orange.
    \label{fig:clearlyflaring_lc}}
\end{figure}

Radio lightcurves of six sources that exhibit high variability are shown in Figure \ref{fig:clearlyflaring_lc}; $5\sigma$ upper limits are adopted for non-detection epochs. These sources illustrate a diverse range of activity, from isolated events to more complex variability. 
DM UMa, which is detected in all available radio epochs (all in VLASS), stands out as the most luminous source in this class, reaching a peak specific luminosity of $1.1\times 10^{18}$ erg s$^{-1}$ Hz$^{-1}$. 
This specific luminosity places DM UMa within the high-luminosity tail of our distribution, while its faintest luminosity is consistent with the peak luminosities in our low-luminosity tail.
In addition, its $V_L$ of 18.1 makes it the source with the third most strongly variable radio emission.
NR Lib exhibits a similarly high peak specific luminosity ($6.8\times 10^{17}$ erg s$^{-1}$ Hz$^{-1}$) in its first epoch, and a high $V_L$ (19.1). Moreover, this $V_L$ is a lower bound on the actual variability as the second epoch is a non-detection. 
Both DM UMa and NR Lib indicate that any interpretation that the radio emission is persistent also would need to allow for a factors of $\gtrsim20$ variability occurring over approximately year-long timescales.

Similarly, sig Gem has high $V_L$ (18.9), but this $V_L$ arises from a strong VLASS detection followed $\sim 10$ days later by a RACS upper-limit. While this could be considered a clear flare-like event, a steep spectral index between RACS and VLASS ($L_\nu \propto \nu^{\alpha}$, where $\alpha>0.5$) could remove sig Gem from the high variability category. Its VLASS-only $V_L$ is a more moderate $\approx5$. In addition, we emphasize that the peak specific luminosity of sig Gem is only $8.9\times 10^{16}$ erg s$^{-1}$ Hz$^{-1}$, about an order of magnitude lower than those of DM UMa or NR Lib.

While HU Vir and V3084 Oph have single-epoch elevated specific luminosities in VLASS ($2.7\times 10^{17}$ erg s$^{-1}$ Hz$^{-1}$ and $1.6\times 10^{17}$ erg s$^{-1}$ Hz$^{-1}$, respectively) bracketed by VLASS non-detections, their peak luminosities are intermediate between DM UMa and sig Gem, and their $V_L$ are lower (13.8 and 10.4, respectively) than DM UMa, NR Lib, and sig Gem. Given the two VLASS non-detections each of HU Vir and V3084 Oph, the measured $V_L$ values here are lower limits on the actual variability of these two sources.

Finally, V0988 Cen has a moderate peak specific luminosity ($2.0\times 10^{17}$ erg s$^{-1}$ Hz$^{-1}$) and an accurate $V_L$ of 11.1 (since it is measured between VLASS detections). In addition, the RACS detection between VLASS Epochs 1 and 2 combines with the RACS non-detection between VLASS Epochs 2 and 3 to create a light curve that could arise from persistent radio emission that increased by factor of $\sim10$ over 3 years before decreasing by a similar factor over 3 years.

There were four sources (V1595 Sco, V0597 Peg, V1362 Tau, IM Peg) among the ``undetermined'' sources that were detected at between 10--20 SNR in a single epoch, but have a $V_L<2$ due to the noise level in undetected epochs. The rest of the ``undetermined'' sources had a maximum SNR $<10$. At these levels, we cannot distinguish whether the detected radio emission arose from a strong upwards variation (including a potential flare), from a more moderate level of upward variability, or from small changes in the image noise. In the latter two cases, the typical level of radio emission in a source would need to be relatively close to, but below, the detection limit.

\subsection{Radio \& X-ray Luminosity Relations} \label{subsec:Lrlx}
The observed correlation between the radio and X-ray luminosities for RS CVn \citep{Drake_1989} and other active stars was developed into an empirical correlation between the specific (5--9 GHz) radio luminosity ($L_{R, \nu}$) and soft (0.5--2.4 keV) X-ray luminosity ($L_{X, {\rm soft}}$), known canonically as the G{\"u}del--Benz (GB) relation \citep{Guedel_1993, Guedel_1995}. Given by $L_{X,soft}/L_{R,\nu} \approx 10^{15.5}$, the relation holds for more than 10 orders of magnitude irrespective of spectral class, age or the level of stellar activity, ranging from active M dwarfs and rapidly rotating FK Comae stars to highly energetic RS CVn and BY Dra binaries. Whether coherent low-frequency ($<3$ GHz) radio emission from stars also follows the GB relation is unclear. \citet{Vedantham_2022} find agreement of their sample with GB, while other samples \citep{Callingham_2021,Driessen_2024,Yiu_2024} find coherent radio emission is typically over-luminous compared to their X-ray luminosity. Regardless, the observed GB relation has been used to link coronal heating, responsible for the broad thermal X-ray emission via flaring activities, to the acceleration of non-thermal electrons responsible for the observed radio emission.

\begin{figure}[t]
    \plotone{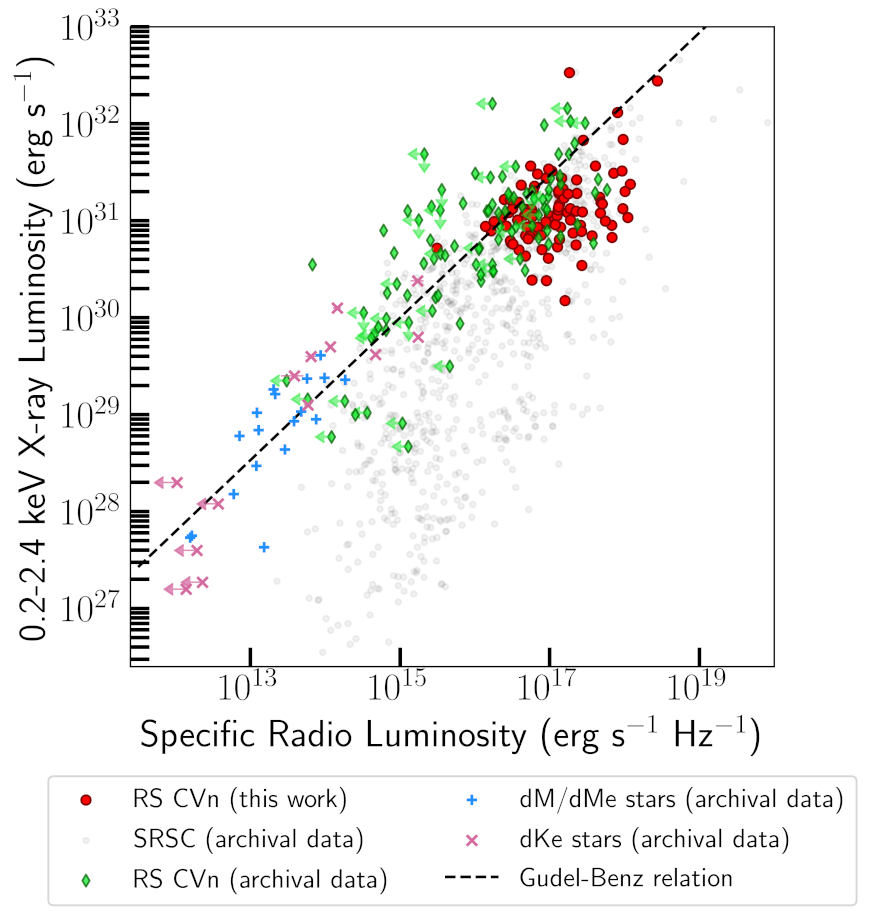}
    \caption{Comparison of the radio and absorption-corrected X-ray luminosities of RS CVn systems to the GB relation. Only the systems with X-ray counterparts are shown here. For each system, only the brightest radio and X-ray detections are plotted. RS CVn candidates are broadly consistent with the GB relation. Previous detections of RS CVn, other flaring stars, and all stars from the SRSC are also shown for reference. RS CVn systems identified here include some of the radio-brightest of this class to ever be detected.
    \label{fig:rscvn_gudelbenz}}
\end{figure}

A comparison of the absorption-corrected X-ray and radio luminosities of our radio-detected sample to a more recent version of the GB relation given by $\log_{10} [L_{R,\nu}] = 1.36$ $(\log_{10}[L_{X,{\rm soft}}] - 18.97)$ \citep{Williams_2014} is shown in Figure \ref{fig:rscvn_gudelbenz}. We only show the brightest radio and X-ray detection for sources detected in multiple epochs. We also plot the brightest detections of RS CVn systems from \citet{Drake_1989}\footnote{Data are taken from \url{https://github.com/AstroLaura/GuedelPlot}}, late (K--M) type stars from archival studies \citep{Gudel_Mdwarf_1993}, and more recent studies \citep{Toet_2021, Vedantham_2022, Yiu_2024} for comparison. Additionally we also include all sources from the published Sydney Radio Star Catalogue \citep[SRSC;][]{Driessen_2024} for reference. The RS CVn candidates we detect in radio and X-rays are consistent with the GB relation and are some of the brightest sources ever detected in this class, with some reaching $L_{R, \nu}$ $\ge 10^{18}$ erg s$^{-1}$ Hz$^{-1}$ in at least one epoch. The detection of these brightest RS CVn candidates is due to the multi-epoch all-sky coverage at higher frequencies compared to previous pointed observations or detections from lower frequency band surveys. The sources we detect appear to be generally radio-bright and/or X-ray dim relative to the GB relation. This could be an artifact resulting from our detection limits; a non-negligible fraction of the single-epoch detections are at the detection threshold, resulting in over-representation of the bright end of the radio luminosities. We also note that that we are only plotting the brightest luminosities of the multi-epoch detections, which could also add to the discrepancy. Furthermore, the non-simultaneous nature of many of the radio and X-ray observations (both in our sample and in past studies), could potentially have X-ray observations capturing the radio-selected sources during non-flaring periods. The $5\sigma$ RASS upper-limits for the 12 radio-detected RS CVn binaries without an X-ray counterpart are consistent with falling slightly below the GB relation, similarly to the rest of our sample. However, because these sources lack confirmed X-ray detections, we refrain from drawing conclusions based solely on their upper-limits.

\begin{figure}[t]
    \plotone{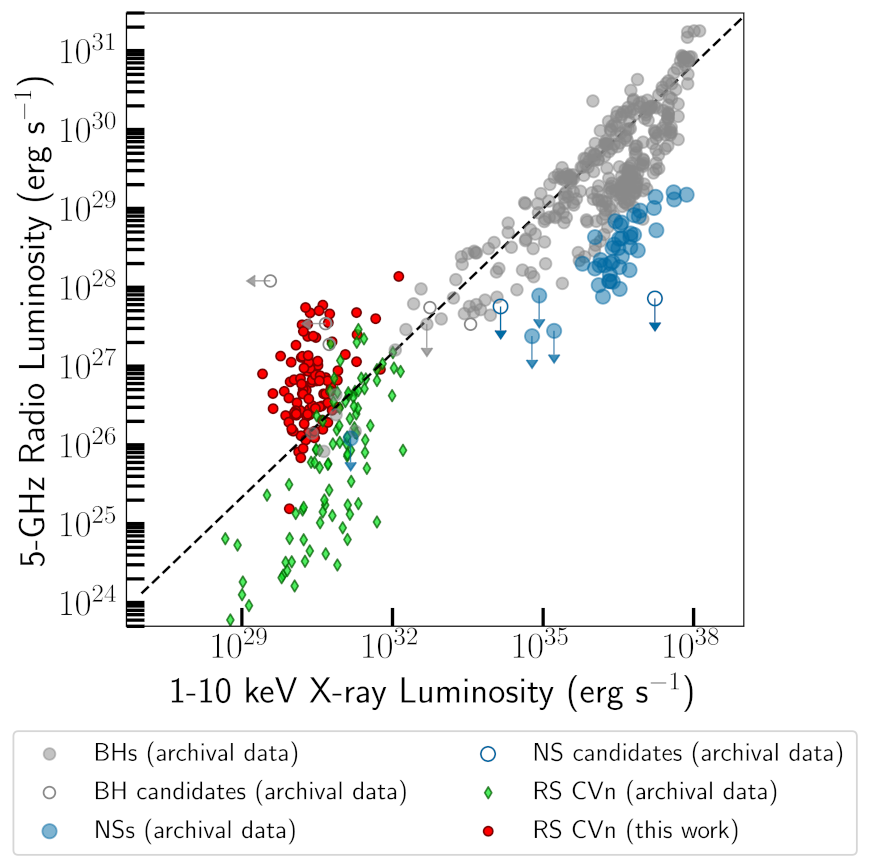}
    \caption{Comparison of the radio and X-ray luminosities of RS CVn to the black hole X-ray binary $L_R-L_X$ relation. The brightest radio and X-ray detections are shown, and only systems with X-ray counterparts are included. The axes are reversed relative to Figure \ref{fig:rscvn_gudelbenz} to follow the convention established in \citet{Gallo_2006}. Other accreting systems such as XRBs, CVs, and previously detected RS CVn are plotted for reference. RS CVn from this work have luminosities comparable to some of the brightest quiescent XRBs ($L_R\sim10^{27} \textrm{–} 10^{28} \rm{\, erg \, s^{-1}}$).
    \label{fig:lrlx}}
\end{figure}

As the detected sources are some of the brightest RS CVn ever detected, we compare them to compact object binaries, which are also known to have a radio--X-ray correlation \citep[e.g.][]{Gallo_2006}. The radio and X-ray luminosities of RS CVn are plotted on the $L_R$--$L_X$ plane for XRBs \citep{arash_bahramian_2022_7059313}, along with the well known relation for black hole XRBs \citep{Gallo_2006} for reference. We assumed a flat radio spectrum when calculating the radio luminosity of the RS CVn candidates at 5~GHz. The radio-bright RS CVn candidates from this work have luminosities comparable to some of the brightest quiescent XRBs ($\ge 10^{28}$ erg s$^{-1}$). \citet{Shishkovsky_2018} reported one such example of a candidate XRB whose properties (e.g. mass function of the visible star of the binary) might be better explained by an ``extreme flaring RS CVn binary''. In the absence of complementary optical or spectroscopic information, bright RS CVn variables could be mistaken for quiescent XRBs. Our sample reinforces that position, and we also raise the issue that some quiescent XRBs may be currently misidentified as RS CVn.

\subsection{Radio Brightness Temperature}\label{subsec:brightnessTemp}
\begin{figure}[t]
    \plotone{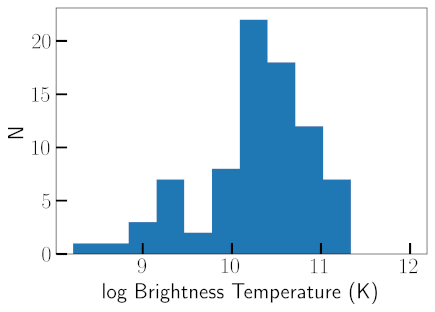}
    \caption{Histogram of the brightness temperature $T_B$ of the radio-detected RS CVn.
    \label{fig:brightness_temp_histogram}}
\end{figure}

To explore the boundaries of emission mechanisms, we estimate the brightness temperature of the radio emission (assuming emission from the whole stellar surface) using the \textit{Gaia} DR3 radius derived by General Stellar Parameterizer from Photometry \citep[\texttt{GSP-Phot};][]{Creevey_2023, Fouesneau_2023}. We assume this value represents the radius of the primary (larger) component. We estimate the  brightness temperature ($T_B$) of the 81 radio-detected sources with valid radius measurements using
\begin{equation}
    T_B \approx 1.38 \times 10^9 \left( \frac{1\textrm{ mas}}{\phi}\right)^{2}\left( \frac{\nu}{1\textrm{ GHz}}\right)^{-2}\left(\frac{S_{\nu}}{1\textrm{ mJy}} \right)\textrm{K},
\end{equation}
where $\phi$ is the angular diameter of the star in milliarcseconds (mas) determined using the \textit{Gaia} DR3 radius, $\nu$ is the radiation frequency in GHz, and $S_{\nu}$ is the specific radio flux density in mJy. The resulting values range between $10^8$ K and $3 \times 10^{11}$ K (as shown in Figure \ref{fig:brightness_temp_histogram}), the upper end of which marks the high end of gyrosynchrotron emission. Brightness temperatures exceeding $10^{12}$ K require either highly Doppler-beamed synchrotron emission, or coherent emission mechanisms \citep{Dulk_1985}. 
Below this limit, as in the case of our sample, one cannot use $T_B$ to distinguish 
between coherent and incoherent emission.

Moreover, since the true radio emission scale is unconstrained, the emission could originate from a scale much larger than the stellar surface, resulting in the actual $T_B$ being lower than reported. Conversely, emission could originate from smaller, localized regions like starspots or magnetic reconnection loops, driving the local $T_B$ past the $10^{12}$ K incoherent limit.
Therefore, while these baseline estimates indicate that both coherent and incoherent mechanisms remain broadly viable across the sample, a definitive characterization is limited without direct constraints on the radio source sizes.

We note that although RS CVn systems are binaries, \texttt{GSP-Phot} assumes a single star model for calculating the radius. This assumption leads to two edge cases: (1) In typical RS CVn systems where the primary star is larger, more massive, more evolved, and more luminous than the secondary star, the reported \textit{Gaia} radius is close to the actual radius of the primary \citep[e.g., FG UMa;][]{Toet_2021}. (2) In the case of twin stellar systems, the reported \textit{Gaia} radius will be $\sim40\%$ larger than the actual radii \citep[e.g., RS CVn itself;][]{Xiang_2020}. However, because $T_B \propto R^{-2}$, a maximum radius overestimation of $\sim40\%$ would have reduced the true $T_B$ by at most $\sim0.3$ dex. Including such uncertainty, all sources still have $T_B<10^{12}$ K (below the incoherent limit) in the case where the radio emitting region arises from the entire surface of the primary.

\section{Discussion} \label{sec:discussion}
\subsection{Lack of Radio Detections of Optically Bright RS CVn} \label{subsec:underDetection}
\begin{figure*}[t]
    \plottwo{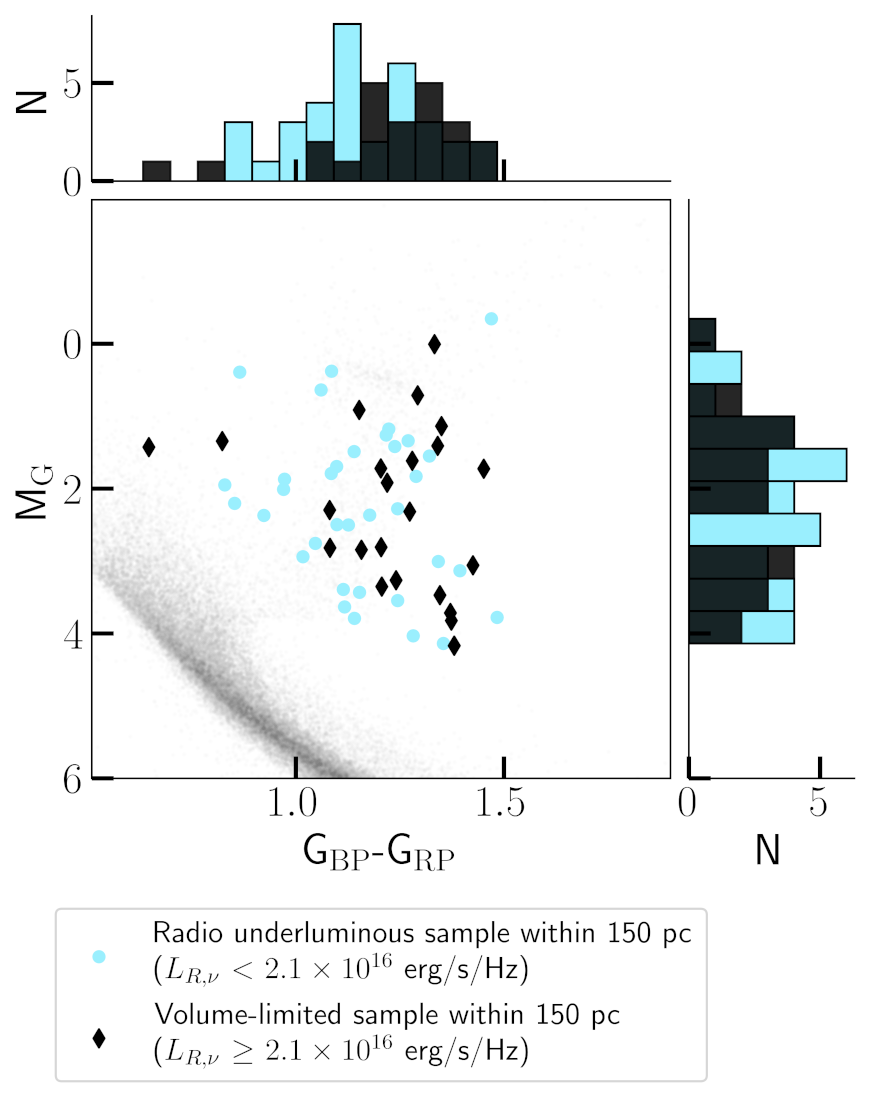}{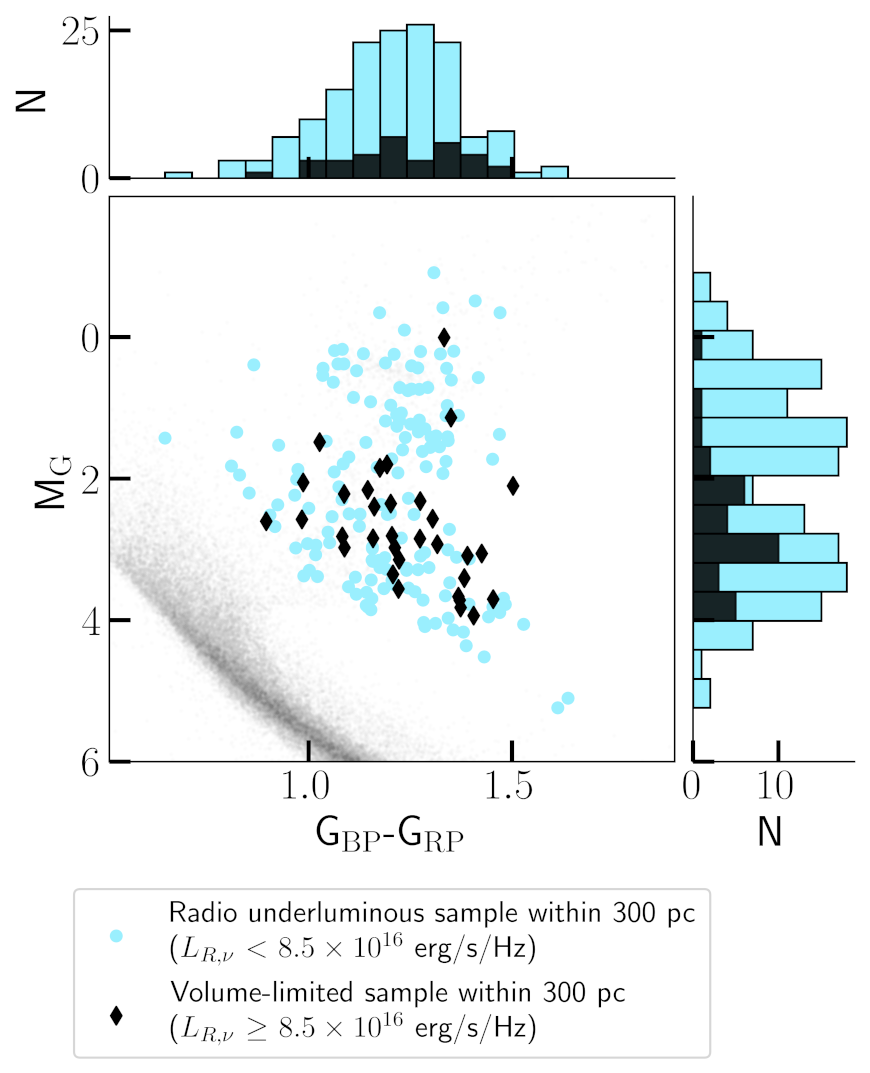}
    \caption{\textit{left}: CMD of the RS CVn binaries within 150 pc. The volume-limited sample of RS CVn candidates with a radio luminosity  L$_{R,\nu} > 2.1 \times 10^{16}$ erg s$^{-1}$ Hz$^{-1}$ are plotted as black diamonds. The RS CVn candidates within this volume below this radio luminosity threshold (including the non-detections) are the radio underluminous sample, and are plotted as cyan circles. The GCNS are shown in grey. \textit{right}: CMD of the RS CVn within 300 pc, with a higher radio luminosity threshold of of $L_{R,\nu} = 8.5 \times 10^{16}$ erg s$^{-1}$ Hz$^{-1}$, with the same symbol scheme. 
    The marginal plots show 1D histograms of $G_{BP}-G_{RP}$ and $M_G$ for the volume-limited radio-bright and radio underluminous samples. Optically brighter RS CVn systems are less likely to have $L_{R,\nu}>8.5\times10^{16}$ erg/s/Hz than optically fainter ones, but they are equally likely to have $L_{R,\nu}>2.1\times10^{16}$ erg/s/Hz as optically fainter ones.
    \label{fig:CMD_volComplete}}
\end{figure*}

Because the RACS detected sample of RS CVn candidates have a lower reliability, we excluded the four sources detected only in RACS from further analysis to ensure robustness. VLASS, RACS, and \textit{Gaia} are all flux-limited surveys. However, \textit{Gaia} is complete out to significantly larger volumes than VLASS or RACS for RS CVn systems. Because \textit{Gaia} is $\sim95\%$ complete to 19th magnitude, our 2-kpc selected VSX--\textit{Gaia} sample --- with a faintest absolute magnitude threshold of $M_G = 5.25$ mag --- is essentially complete, except in regions of strong reddening ($A_V \gtrsim 1.2$). Therefore, to avoid biases that are mostly introduced by radio incompleteness with distance (due to flux density sensitivity), we constructed two volume-limited subsamples: (1) within 150 pc with $L_{R,\nu}$ $>2.1 \times 10^{16} {\rm \, erg \, s^{-1} \, Hz^{-1}}$, chosen to maximize source coverage across the CMD; and (2) within 300 pc with $L_{R,\nu}$ $>8.5 \times 10^{16} {\rm \, erg \, s^{-1} \, Hz^{-1}}$, chosen to maximize the overall sample size. 

Under the assumption that the radio emission is associated with stellar flares, we note that our luminosity and distance limits are only valid for the sources flaring during the time of radio observation.
These luminosity thresholds were set at $6.6\sigma$ (corresponding to a flux density limit of 790 $\mu$Jy --- given a characteristic VLASS image rms of 120 $\mu$Jy); ensuring that 95\% of the sources exceed the 5-$\sigma $ threshold we use to establish a detection (assuming a Gaussian distribution). Figure \ref{fig:CMD_volComplete} shows the CMDs of the two subsamples along with marginal 1-dimensional (1D) histograms of color and magnitude. The volume-limited sources above the appropriate $L_R$ threshold are shown as black diamonds. For sources within the distance limit of each sample, cyan circles indicate the RS CVn candidates with radio luminosity below the luminosity threshold for that sample (whether detected or not); the `radio underluminous sample'.

To determine if the volume-limited and radio underluminous samples are biased by distance, we performed a 1D two-sample bootstrapped Kolmogorov–Smirnov (KS) test on the distance distributions. These analyses showed no statistically significant difference in either the 150 pc or the 300 pc volume-limited samples. Similar 1D and two-dimensional (2D) bootstrapped KS tests on the $G_{BP}-G_{RP}$ color and the CMD, respectively, also revealed no significant differences, as evident in Figure \ref{fig:CMD_volComplete} and the marginal histograms. 

The $M_G$ histogram shows a uniform detection fraction across $M_G$ for the 150 pc sample, which has a lower radio luminosity threshold. In contrast, the 300 pc sample, with a higher luminosity threshold, exhibits a clear under-detection of optically bright RS CVn ($M_G \leq 2$, hereafter \textit{b}-RS) relative to the fainter population ($M_G > 2$, hereafter \textit{f}-RS). This difference is statistically significant, as confirmed by a bootstrapped KS test ($p = 0.008$). The \textit{f}-RS are approximately five times more likely to be above the higher radio threshold than the \textit{b}-RS, with their detection-fraction confidence intervals overlapping only at $>3\sigma$, signifying a statistically significant difference. The difference in detection rate between the 150 pc and the 300 pc samples suggests that \textit{b}-RS are generally less likely to be radio-loud than \textit{f}-RS.

Here we would like to note a critical caveat; $\sim90\%$ of the VSX RS CVn catalog and $\sim80\%$ of our parent $2\text{ kpc}$ VSX-\textit{Gaia} catalog consist of sources from the Zwicky Transient Facility \citep[ZTF;][]{Chen_2020}. Because the ZTF's automated classification algorithm requires high SNR of both flux and periodic modulation, optically fainter systems --- primarily \textit{f}-RS  --- are more likely to be missed, especially at larger distances. Conversely, ZTF's saturation limit ($\sim12-13\text{ mag}$) means brighter nearby systems (bright \textit{b}-RS) may also be missed. The compilation of diverse surveys and studies that constitute the VSX catalog makes it difficult to perform an in-depth analysis of all such underlying heterogeneous selection biases, and thus beyond the scope of this paper. Nonetheless, such incompleteness could artificially produce/enhance our observed trends. Nevertheless, because no other catalog of RS CVn systems exists at this scale, the VSX catalog represents the most comprehensive collection of known RS CVn systems. Therefore, the trends observed above and in subsequent sections strictly reflect the properties of the currently known VSX-\textit{Gaia} RS CVn population.

\subsection{Stellar Activity} \label{subsec:stellarActivity}

\begin{figure*}[t]
    \plotone{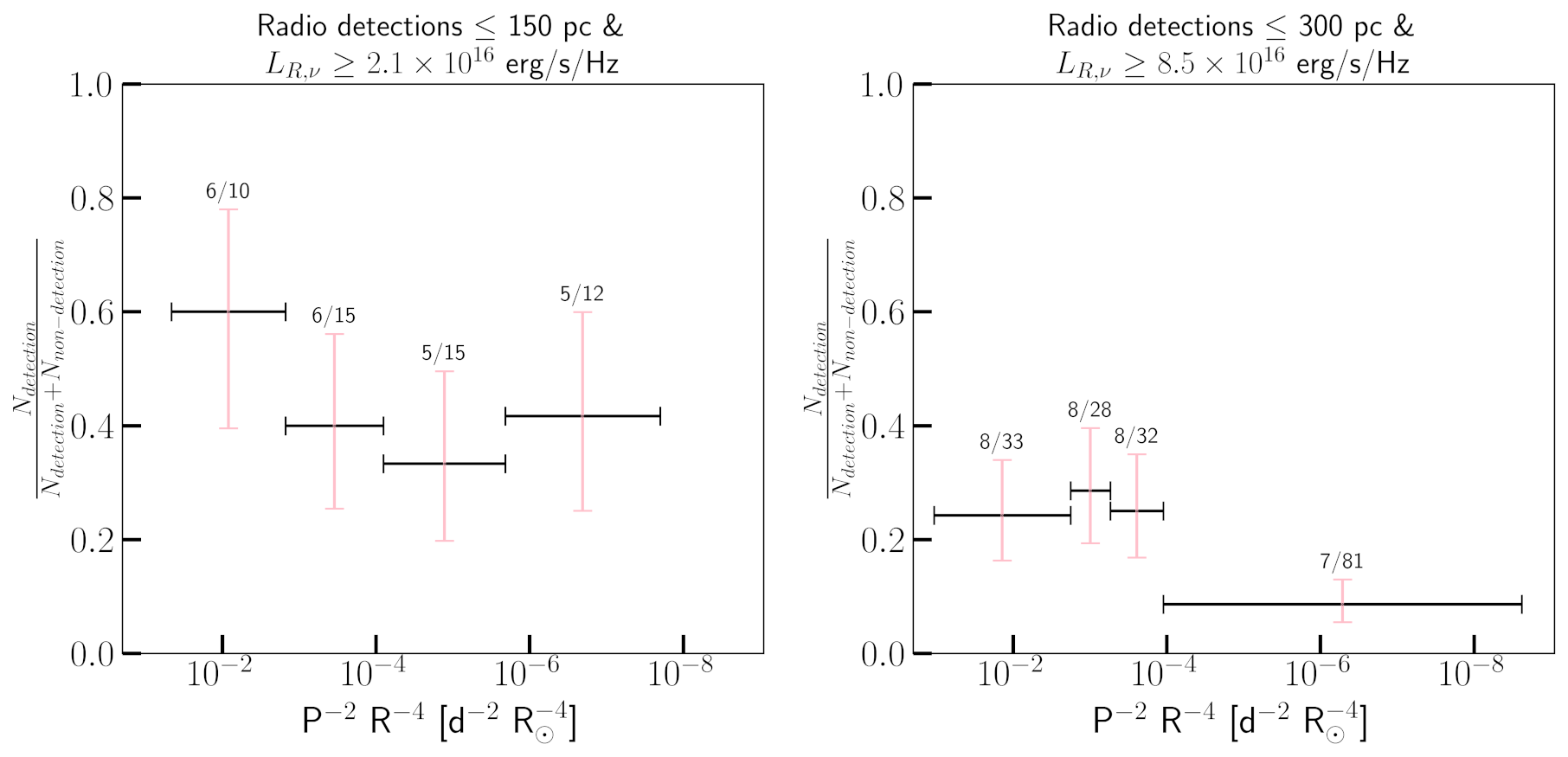}
    \caption{Radio-detection fraction as a function of $P^{-2}R^{-4}$ \citep[$\approx \frac{1}{\mathrm{R}_\mathrm{o}^2}$;][]{Reiners_2014} for both volume-limited samples. The X-axis is inverted so that $\mathrm{R}_\mathrm{o}$ increases to the right, to be consistent with past work \citep[e.g.,][]{Reiners_2014}. Each bin contains an equal number of sources from a volume-limited sample; the fraction is given above each bins.  The bin widths are shown in black and detection-fraction errors in pink. Although the error bars are large, the detection fraction is consistent with being highest at the largest $P^{-2}R^{-4}$ values. This is consistent with radio emission being associated with stellar activity. \label{fig:stellarActivityHist}}
    
\end{figure*}

The magnetic field generation in late-type stars with convective envelopes is usually explained with the $\alpha\omega$-dynamo model \citep{Parker_1955}, which attributes the generation and sustenance of the complex magnetic fields to turbulent convection flow ($\alpha$ effect) and differential rotation ($\omega$ effect). It is quantified by a dimensionless parameter, the Rossby number, $\mathrm{R}_\mathrm{o}$ $= P_s / \tau_c$, where $P_s$ is the rotation period and $\tau_c$ the convective overturn time. Studies have shown magnetic activity, as traced by indicators such as Ca II, H$\alpha$ and X-ray emission, increases with decreasing $\mathrm{R}_\mathrm{o}$ \citep{Noyes_1984, Pizzolato_2003}. 

Following \citet{Reiners_2014}, we adopted $P^{-2}R^{-4}$ as a proxy for stellar activity ($\approx \mathrm{R}_\mathrm{o}^{-2}$) for our analysis. Here $P$ is the VSX-reported period, generally the photometrically derived period; we attribute such periodicity to the spin and make the standard assumption that the spin period and the orbital period have synchronized due to tidal effects. $R$ is the \texttt{GSP-Phot} derived \textit{Gaia} radius \citep{Creevey_2023, Fouesneau_2023}, which we assume to be the radius of the primary. While the relation given by \citet{Reiners_2014} was originally calibrated for low-mass dwarfs ($< 1.4$ M$_\odot$), it was verified to match the classical relations suggested by \citet{Pallavicini_1981} and \citet{Pizzolato_2003}. Both of these studies base their work on the even earlier discovery of a similar activity–rotation relation found across a wider mass range for RS CVn systems by \citet{Walter_1981}, which closely matches the \citet{Pizzolato_2003} trends. This empirical agreement between low-mass dwarfs and RS CVn systems holds as deep convective envelopes are present in both types, resulting in the same underlying physics being responsible for their stellar activity. 

To statistically evaluate whether radio-detection is dependent on this stellar activity proxy, we performed a 1D bootstrapped KS on the $P^{-2}R^{-4}$ distributions of the radio-detected and radio-underluminous samples. For the 150 pc volume limited sample, the test shows no statistically significant evidence of two distinct populations. However, the test of the 300 pc sample clearly demonstrates that the radio-detected and radio-underluminous distributions are drawn from fundamentally different parent populations ($p = 0.004$). This is evidenced by the radio-detection fraction as a function of $P^{-2}R^{-4}$ shown in Figure \ref{fig:stellarActivityHist} for both volume-limited samples. The detection fractions are computed in bins of $P^{-2}R^{-4}$ such that each bin contains an approximately equal number of sources from the radio-detected volume-limited samples. The black bars indicate the bin widths, while the pink error bars represent 1-$\sigma$ Gehrels uncertainties \citep[][assuming a binomial distribution]{Gehrels_1986} on the detection fractions. The fractions of the radio-detected and the radio-underluminous samples are given above each bin for reference. Although the error bars are large (due to small number statistics), the 300 pc volume-limited sample shows higher detection fractions at $P^{-2}R^{-4} \gtrsim 10^{-4} {\rm \, d^{-2} R_\odot^{-4}}$, corresponding to lower $\mathrm{R}_\mathrm{o}$ and hence higher stellar activity. This region of the histogram, which is preferentially radio-loud by virtue of meeting our detection thresholds, is dominated by the \textit{f}-RS with the highest stellar activity. 
Given the poor statistical constraints in the 150 pc sample, we refrain from drawing a conclusion for this more restricted sample.

\begin{figure*}[t]
    \plotone{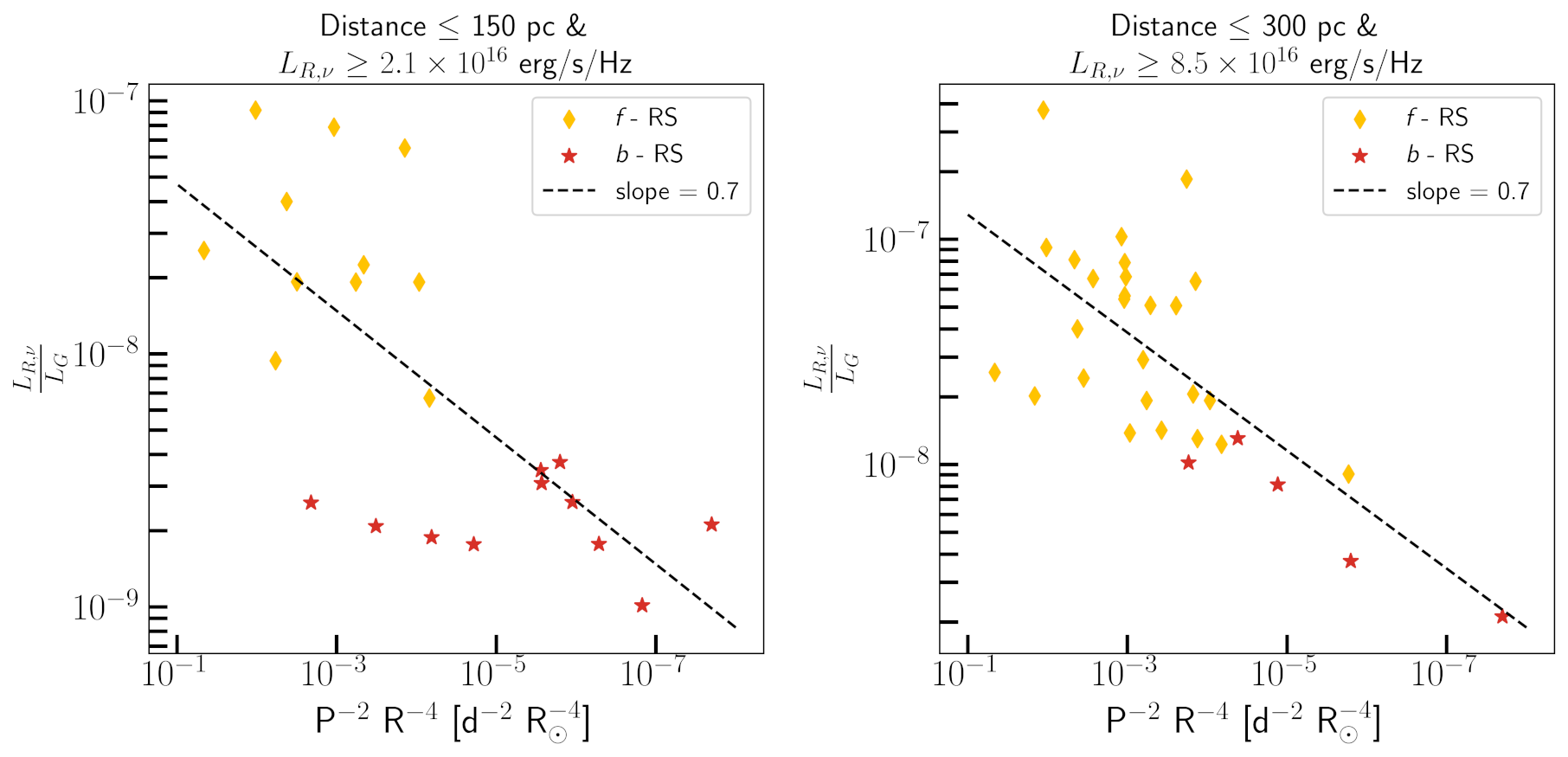}
    \caption{Radio-to-optical luminosity fraction as a function of $P^{-2}R^{-4}$ \citep[$\approx \frac{1}{\mathrm{R}_\mathrm{o}^2}$;][]{Reiners_2014} for both volume-limited samples. The X-axis is inverted so that $\mathrm{R}_\mathrm{o}$ increases to the right. This emphasizes that \textit{f}-RS radiate a larger fraction of their luminosity in the radio compared to \textit{b}-RS, which is consistent with a link between radio emission and  stellar activity.
    \label{fig:stellarActivity}}
\end{figure*}

The observed radio-loudness in the region dominated by the \textit{f}-RS is further illustrated in the relation between radio-to-optical luminosity fraction and $P^{-2}R^{-4}$, as shown for both volume-limited samples in Figure \ref{fig:stellarActivity}. We adopt the radio-to-optical luminosity as the total radio luminosity fraction, since bolometric correction is not available for all \textit{Gaia} sources to calculate the bolometric luminosity. This approximation is justified by our analysis of the available subsample, where we find that the bolometric luminosity deviates from the multi-scan averaged \textit{Gaia} $G$-band luminosity by less than $6\%$. \textit{f}-RS sources at lower $P^{-2}R^{-4}$ exhibit higher radio luminosity fractions, indicative of enhanced stellar activity. If the radio luminosity scaled linearly with the bolometric luminosity, the fractional radio luminosity would remain invariant across $P^{-2}R^{-4}$. Instead, \textit{f}-RS sources emit a larger fraction of their total luminosity in the radio. Considering most RS CVn systems are tidally synchronized, the shorter periods and small radii seen in these fractionally more radio luminous \textit{f}-RS sources hints towards boosted magnetic activity. In contrast, the radio luminosity fraction seen in \textit{b}-RS --- that are mainly on the red giant branch, and have longer periods and larger radii --- suggests that RS CVn with at least one giant component experience reduced stellar activity, making them less likely to be radio-loud.
It remains unclear, however, whether the enhanced activity in \textit{f}-RS results primarily from more frequent flares, more energetic flares, or a combination of both. More extensive samples, both in wider temporal coverage and lower luminosity limits, are needed to address the origin of enhanced activity in \textit{f}-RS.

A notable caveat in adapting the $P^{-2}R^{-4}$ stellar activity proxy is its $R^{-4}$ dependency on the \textit{Gaia} radius, which can overestimate the primary radius (as detailed in Section \ref{subsec:brightnessTemp}). This overestimate can reduce the true $P^{-2}R^{-4}$ value by at most $\sim0.6$ dex. However, this uncertainty does not alter either of the main conclusions discussed above. The histogram in Figure \ref{fig:stellarActivityHist} is constructed using roughly equal sources per bin. This overestimation will primarily affect the individual bin widths rather than the source counts. In Figure \ref{fig:stellarActivity}, although this uncertainty might change the individual position of the sources on this plot, it will do so only along the horizontal $P^{-2}R^{-4}$ axis, leaving our overall conclusion about the radio luminosity fraction of the \textit{f}-RS and \textit{b}-RS unaffected.

\subsection{Persistent Radio-Bright Emitters} \label{subsec:persistence}
Among the ``strongly varying'' subclass of the radio-detected RS CVn candidates discussed in Section \ref{subsec:measuredVar}, two of them --- ASAS J060415+1245.9 (HD 251108) and V0340 Gem --- showed persistent and high radio luminosities of $\ge 2 \times 10^{17}$ erg s$^{-1}$ Hz$^{-1}$ across all available epochs (5 and 3 epochs, respectively); these detections have SNR from 6--77 and 12--26, respectively. These detections in all available epochs suggest that their emission is unlikely to be from infrequent stellar flares. Persistent, yet variable, emission appears to be more likely. 

ASAS J060415+1245.9 (HD 251108), located at a distance of $\sim$514 pc \citep{Gaia_2023}, was detected in all 3 VLASS epochs and 2 RACS-low epochs with a radio luminosity range of 1.1--2.8 $\times 10^{18} {\rm \, erg \, s^{-1} \, Hz^{-1}}$. V0340 Gem, at $\sim$467 pc \citep{Gaia_2023}, was detected in all three epochs of VLASS with a luminosity range of 4.0--9.5 $\times 10^{17} {\rm \, erg \, s^{-1} \, Hz^{-1}}$ in the 2--4 GHz band. These two sources are the brightest RS CVn candidates known to date that have remained persistently bright over a period of at least seven years, spanning the VLASS and RACS-low observation epochs. However, their sustained and relatively constant flux is difficult to explain using flare-based emission models for RS CVn. 
Although persistent radio emission is more likely, this raises an open question as to why these two sources are persistently radio emitting above $10^{18} {\rm \, erg \, s^{-1} \, Hz^{-1}}$.

Such persistent radio-brightness is characteristic of XRBs. However, both HD 251108 and V0340 Gem are relatively well studied sources with little ambiguity on their non-XRB source type \citep[e.g.,][]{Erdem_2009, Halbwachs_2014, Fuhrmeister_2025, Mao_2025}. But, in the absence of complementary multi-wavelength information, such persistently radio-bright sources might be misclassified as XRBs, particularly when the radio and X-ray luminosities of these systems are comparable to V404 Cyg, one of the brightest quiescent XRBs.

\subsection{Caveats} \label{subsec:caveats}
\begin{figure*}[t]
    \plotone{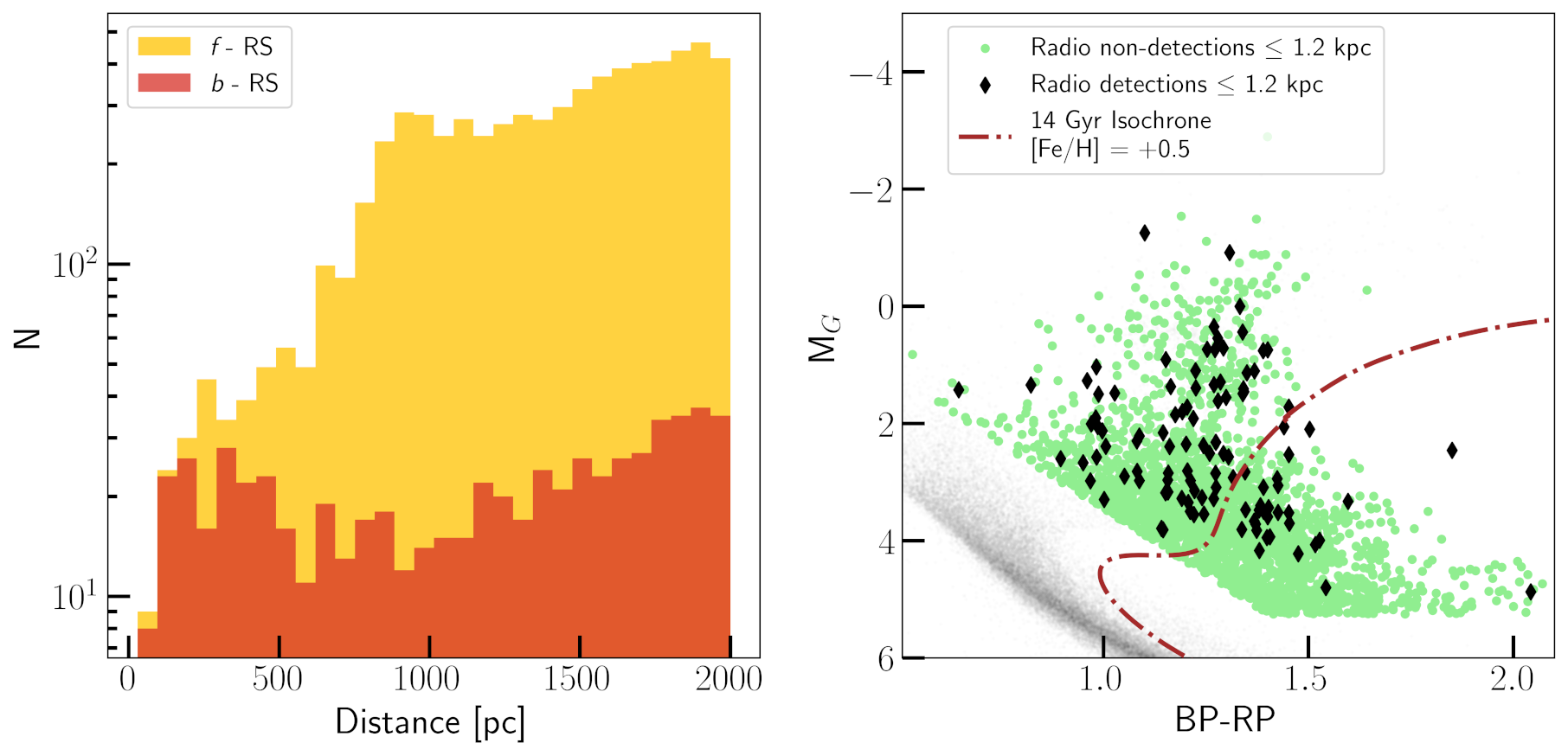}
    \caption{\textit{left}: Histograms of the distances to the \textit{b}-RS and \textit{f}-RS within 900 pc. The number of \textit{b}-RS sources decreases beyond 300 pc before rising again at around 1 kpc, as opposed to increasing monotonically with distance as expected and seen for \textit{f}-RS sources. \textit{right}: CMD of RS CVn candidates within 900 pc, with radio-detections shown as black diamonds and non-detections shown in green. The 14 Gyr, [Fe/H] = +0.5 MIST isochrone is shown as the brown dot-dashed line. Radio detections appear to be rarer along the Horizontal Branch ($1\gtrsim M_G \gtrsim  0$) and in the sub-subgiant region (region redder than the isochrone).    
    \label{fig:caveats}}
\end{figure*}

In a volume-complete sample, differential source counts are expected to increase with distance. However, our VSX–\textit{Gaia} catalog shows a decline in the number of \textit{b}-RS systems between $\sim$300--1000 pc, whereas \textit{f}-RS source counts steadily increase out to 1 kpc (see Figure \ref{fig:caveats}, left panel). This result does not appear to be due to our inclusion of high extinction regions; the same pattern is seen in the RS CVn catalog from \citet{Leiner_2022} that avoids $A_V>0.5$. Moreover, if extinction was a primary effect, one would expect problems with \textit{f}-RS systems and not \textit{b}-RS systems. Since both our catalog and that from \citet{Leiner_2022} start with VSX cross-matched with \textit{Gaia}, there could be currently unidentified selection criteria in the parent sample. A full detailed statistical analysis of this discrepancy and its origin is beyond the scope of this paper.

The flux density-limited nature of the radio surveys likely restricts our radio-detected sample to 1.2 kpc, since only the intrinsically luminous sources are detected at larger distances. Within our sample we see three areas on the CMD (Figure \ref{fig:caveats}, right panel) that have fewer RS CVn radio detections: (1) along the Horizontal Branch ($1\gtrsim M_G \gtrsim  0$); (2) in the region close to the main sequence (e.g., $B_P-R_P \lesssim 1$ at $M_G \sim 3$ through $B_P-R_P \lesssim 1.5$ at $M_G \gtrsim 4$); and (3) in the region redder than the 14 Gyr, [Fe/H] = +0.5 MIST  isochrone \citep{Dotter_2016}, which marks the reddest possible edge of the giant branch, representing the location of the oldest and most metal-rich population that could possibly exist, and therefore marks the upper edge of the sub-subgiant region. We caution that in all three cases, our sample size limits our ability to determine if these results are statistically significant, let alone to probe the origin of these deficits if real. Moreover, these results may arise from unidentified selection criteria. Nevertheless, we identify these regions for potential followup in future surveys.

\section{Conclusion} \label{sec:conclusion}
We performed a systematic search for RS CVn systems using the VSX catalog and identified a total of 108 candidates with radio counterparts in the VLASS and RACS all-sky radio surveys. We cross-matched the detected RS CVn with the eRASS1 and 2RXS catalogues to find their X-ray counterparts. Over half the sample exhibits significant, high-SNR variability ($\geq 2$), with six of them showing clear signs of flaring events. This confirmed that these RS CVn candidates are dominated by intrinsic, transient radio emission rather than image noise. These radio-detected RS CVn systems include some of the brightest radio sources of this class observed to date. Notably, some of the RS CVn candidates have radio and X-ray luminosities comparable to quiescent black hole X-ray binaries, highlighting  potential misclassifications between the two classes. 

The brightness temperature for all of the radio-detected RS CVn systems with valid \textit{Gaia} radius estimates is below $10^{12}$ K, implying we cannot clearly identify whether radio emission from our sources arises from a coherent or incoherent emission process.

Comparison of the detection fraction of optically brighter and fainter RS CVn systems in two different volume-limited samples suggests that RS CVn binaries with at least one giant component are less likely to be radio-loud, compared to their optically fainter counterparts. Additionally, the optically fainter systems also tend to have higher $P^{-2}R^{-4}$ (lower $\mathrm{R}_\mathrm{o}$) and larger radio luminosity fraction, consistent with higher stellar activity --- possibly due to stronger tidal synchronization arising from shorter periods and smaller radii. 

However, among the 108 detected RS CVn binaries, two optically bright systems --- ASAS J060415+1245.9 (HD 251108) and V0340 Gem --- show signs of persistent and high radio luminosities ($\ge 10^{27}$ erg s$^{-1}$ at 5~GHz) across all epochs. This persistent brightness suggests that these systems may be fundamentally different from typical RS CVn binaries and warrants further investigation.

\begin{acknowledgments}
SSR and GRS are supported by NSERC Discovery Grant RGPIN-2021-04001.
COH is supported by NSERC Discovery Grant RGPIN-2023-04264, and Alberta Innovates Advance Program 242506334. DK was supported by NSF grant AST-2511757.
This scientific work uses data obtained from Inyarrimanha Ilgari Bundara, the CSIRO Murchison Radio-astronomy Observatory. We acknowledge the Wajarri Yamaji People as the Traditional Owners and native title holders of the Observatory site. CSIRO’s ASKAP radio telescope is part of the Australia Telescope National Facility (https://ror.org/05qajvd42). Operation of ASKAP is funded by the Australian Government with support from the National Collaborative Research Infrastructure Strategy. ASKAP uses the resources of the Pawsey Supercomputing Research Centre. Establishment of ASKAP, Inyarrimanha Ilgari Bundara, the CSIRO Murchison Radio-astronomy Observatory and the Pawsey Supercomputing Research Centre are initiatives of the Australian Government, with support from the Government of Western Australia and the Science and Industry Endowment Fund.
The National Radio Astronomy Observatory is a facility of the U.S. National Science Foundation operated under cooperative agreement by Associated Universities, Inc.
This research has made use of the VizieR catalogue access tool, CDS, Strasbourg, France \citep{10.26093/cds/vizier}. The original description of the VizieR service was published in \citet{vizier2000}.
This work is based on data from eROSITA, the soft X-ray instrument aboard SRG, a joint Russian-German science mission supported by the Russian Space Agency (Roskosmos), in the interests of the Russian Academy of Sciences represented by its Space Research Institute (IKI), and the Deutsches Zentrum f$\ddot{\textrm{u}}$r Luft- und Raumfahrt (DLR). The SRG spacecraft was built by Lavochkin Association (NPOL) and its subcontractors, and is operated by NPOL with support from the Max Planck Institute for Extraterrestrial Physics (MPE). The development and construction of the eROSITA X-ray instrument was led by MPE, with contributions from the Dr. Karl Remeis Observatory Bamberg \& ECAP (FAU Erlangen-Nuernberg), the University of Hamburg Observatory, the Leibniz Institute for Astrophysics Potsdam (AIP), and the Institute for Astronomy and Astrophysics of the University of T$\ddot{\textrm{u}}$bingen, with the support of DLR and the Max Planck Society. The Argelander Institute for Astronomy of the University of Bonn and the Ludwig Maximilians Universit$\ddot{\textrm{a}}$t Munich also participated in the science preparation for eROSITA.
This research has made use of the International Variable Star Index (VSX) database, operated at AAVSO, Cambridge, Massachusetts, USA.
This work presents results from the European Space Agency (ESA) space mission \textit{Gaia}. \textit{Gaia} data are being processed by the Gaia Data Processing and Analysis Consortium (DPAC). Funding for the DPAC is provided by national institutions, in particular the institutions participating in the Gaia MultiLateral Agreement (MLA).
SSR, GRS, and COH are members of the University of Alberta. The University of Alberta, its buildings, labs and research stations are primarily located on the territory of N\'ehiyaw (Cree), Niitsitapi (Blackfoot), M\'etis, Nakoda (Stoney), Dene, Haudenosaunee (Iroquois) and Anishinaabe (Ojibway/Saulteaux), lands that are now known as part of Treaties 6, 7 and 8 and homeland of the M\'etis. The University of Alberta and its members respect the sovereignty, lands, histories, languages, knowledge systems and cultures of all First Nations, M\'etis, and Inuit.

\end{acknowledgments}

\software{Astropy \citep{astropy:2013, astropy:2018, astropy:2022}, 
          Astroquery \citep{Astroquery:2019},
          TOPCAT \citep{Topcat:2005},
          NumPy \citep{numpy:2020},
          Pandas \citep{pandas:2020},
          SciPy \citep{SciPy:2020},
          Matplotlib \citep{matplotlib:2007},
          seaborn \citep{seaborn:2021}
          }

\bibliography{mainbib}{}
\bibliographystyle{aasjournalv7}

\begin{deluxetable*}{clcr}
\tablecaption{A descriptive table of the final RS CVn candidate catalog used for this study. 
\label{tab:allRSCVn_MR}}
\tablehead{
   \colhead{Row number} & \colhead{Units} & \colhead{Label} & \colhead{Description}
}
\startdata
1 & \nodata & RSHindex & Source identifier \\
2 & \nodata & Name & VSX source name \\
3 & deg & RAdeg & Gaia DR3 Right Ascension (J2000) at Ep=2016.0 \\
4 & deg & DEdeg & Gaia DR3 Declination (J2000) at Ep=2016.0 \\
5 & mas/yr & pmRA & Gaia DR3 proper motion in right ascension \\
6 & mas/yr & e\_pmRA & Gaia DR3 standard error of proper motion in right ascension \\
7 & mas/yr & pmDE & Gaia DR3 proper motion in declination \\
8 & mas/yr & e\_pmDE & Gaia DR3 standard error of proper motion in declination \\
9 & pc & Dist & Gaia DR3 distance \\
10 & pc & e\_Dist & Gaia DR3 standard error of distance \\
11 & mag & Gmag & Gaia DR3 G-band magnitude (extinction corrected) \\
12 & mag & BP-RP & Gaia DR3 BP-RP colour (extinction corrected) \\
13 & mag & AV & V-band extinction \\
14 & R$_\odot$ & Rad & Gaia DR3 radius estimation \\
15 & day & Period & VSX reported period \\
16 & \nodata & RadioDetectedFlag & Flag indicates this work detects a radio source \\
\enddata
\tablecomments{This table is published in its entirety in the electronic edition of the {\it Astrophysical Journal}.  The form of it is shown here for guidance.}
\end{deluxetable*}

\begin{deluxetable*}{cccc}
\tablecaption{A descriptive table of the radio and X-ray properties of the radio-detected RS CVn systems.
\label{tab:validRSCVn_MR}}
\tablehead{
   \colhead{Row number} & \colhead{Units} & \colhead{Label} & \colhead{Description}
}
\startdata
1 & \nodata & RSHindex & Source identifier \\
2 & deg & radioRAdeg & Right Ascension (J2000) from brightest radio detection; epoch listed under radioEpoch \\
3 & deg & radioDEdeg & Declination (J2000) from brightest radio detection; epoch listed under radioEpoch \\
4 & \nodata & radioEpoch & Survey epoch of the brightest radio detection \\
5 & yyyy-mm-dd & radioObsDate & Observation date of the brighest radio detection; epoch list under radioEpoch \\
6 & arcsec & radioOpticalSep & Positional offset between radio position and \textit{Gaia} DR3 position \\
7 & mJy & VLASS-E1-FluxDens & Flux density from VLASS Epoch 1 \\
8 & mJy & e\_VLASS-E1-FluxDens & Standard error in flux density from VLASS Epoch 1 \\
9 & 10$^{-7}$ W Hz$^{-1}$ & VLASS-E1-L & Specific luminosity from VLASS Epoch 1 (erg s$^{-1}$ Hz$^{-1}$)\\
10 & 10$^{-7}$ W Hz$^{-1}$ & e\_VLASS-E1-L & Standard error in specific luminosity from VLASS Epoch 1 (erg s$^{-1}$ Hz$^{-1}$)\\
11 & mJy & VLASS-E2-FluxDens & Flux density from VLASS Epoch 2 \\
12 & mJy & e\_VLASS-E2-FluxDens & Standard error in flux density from VLASS Epoch 2 \\
13 & 10$^{-7}$ W Hz$^{-1}$ & VLASS-E2-L & Specific luminosity from VLASS Epoch 2 (erg s$^{-1}$ Hz$^{-1}$)\\
14 & 10$^{-7}$ W Hz$^{-1}$ & e\_VLASS-E2-L & Standard error in specific luminosity from VLASS Epoch 2 (erg s$^{-1}$ Hz$^{-1}$)\\
15 & mJy & VLASS-E3-FluxDens & Flux density from VLASS Epoch 3 \\
16 & mJy & e\_VLASS-E3-FluxDens & Standard error in flux density from VLASS Epoch 3 \\
17 & 10$^{-7}$ W Hz$^{-1}$ & VLASS-E3-L & Specific luminosity from VLASS Epoch 3 (erg s$^{-1}$ Hz$^{-1}$)\\
18 & 10$^{-7}$ W Hz$^{-1}$ & e\_VLASS-E3-L & Standard error in specific luminosity from VLASS Epoch 3 (erg s$^{-1}$ Hz$^{-1}$)\\
19 & mJy & RACS-E1-FluxDens & Flux density from RACS Epoch 1 \\
20 & mJy & e\_RACS-E1-FluxDens & Standard error in flux density from RACS Epoch 1 \\
21 & 10$^{-7}$ W Hz$^{-1}$ & RACS-E1-L & Specific luminosity from RACS Epoch 1 (erg s$^{-1}$ Hz$^{-1}$)\\
22 & 10$^{-7}$ W Hz$^{-1}$ & e\_RACS-E1-L & Standard error in specific luminosity from RACS Epoch 1 (erg s$^{-1}$ Hz$^{-1}$)\\
23 & mJy & RACS-E2-FluxDens & Flux density from RACS Epoch 2 \\
24 & mJy & e\_RACS-E1-FluxDens & Standard error in flux density from RACS Epoch 2 \\
25 & 10$^{-7}$ W Hz$^{-1}$ & RACS-E2-L & Specific luminosity from RACS Epoch 2 (erg s$^{-1}$ Hz$^{-1}$)\\
26 & 10$^{-7}$ W Hz$^{-1}$ & e\_RACS-E2-L & Standard error in specific luminosity from RACS Epoch 2 (erg s$^{-1}$ Hz$^{-1}$)\\
27 & mW m$^{-2}$ & eRASS1-softFlux & Flux in the (0.2--2.3 keV) soft band from eRASS1 (erg s$^{-1}$ cm$^{-2}$)\\
28 & 10$^{-7}$ W & eRASS1-softL & Unabsorbed (corrected using the derived NH) luminosity in the (0.2--2.3 keV) soft band from eRASS1 (erg s$^{-1}$)\\
29 & mW m$^{-2}$ & eRASS1-hardFlux & Flux in the (2.3--5.0 keV) hard band from eRASS1 (erg s$^{-1}$ cm$^{-2}$)\\
30 & 10$^{-7}$ W & eRASS1-hardL & Unabsorbed (corrected using the derived NH) luminosity in the (2.3--5.0 keV) hard band from eRASS1 (erg s$^{-1}$)\\
31 & mW m$^{-2}$ & 2RXS-softFlux & Flux in the (0.2--2.3 keV) soft band from 2RXS (erg s$^{-1}$ cm$^{-2}$)\\
32 & 10$^{-7}$ W & 2RXS-softL & Unabsorbed (corrected using the derived NH) luminosity in the (0.2--2.3 keV) soft band from 2RXS (erg s$^{-1}$)\\
33 & mW m$^{-2}$ & 2RXS-hardFlux & Flux in the (2.3--5.0 keV) hard band from 2RXS (erg s$^{-1}$ cm$^{-2}$)\\
34 & 10$^{-7}$ W & 2RXS-hardL & Unabsorbed (corrected using the derived NH) luminosity in the (2.3--5.0 keV) hard band from 2RXS (erg s$^{-1}$)\\
35 & cm$^{-2}$ & NH & Hydrogen column density derived from AV \\
36 & ... & VariabilityIndex & Ratio of the maximum radio luminosity to the minimum radio luminosity (including non-detection upper-limits) \\
\enddata
\tablecomments{This table is published in its entirety in the electronic edition of the {\it Astrophysical Journal}.  The form of it is shown here for guidance.}
\end{deluxetable*}

\begin{longrotatetable}
\movetabledown=20mm
\begin{deluxetable*}{clccccccccccccc}
\tabletypesize{\footnotesize}
\tablecaption{Optical, radio and X-ray properties, along with the VSX-reported period and primary radius estimates of the radio detected RS CVn candidates.
\label{tab:validRSCVn_radioXray}}
\tablehead{
\colhead{RSH Index} & \colhead{VSX Name} & \colhead{RAdeg} & \colhead{DEdeg} & \colhead{Dist.} & \colhead{M$_G$} & \colhead{$G_{BP}-G_{RP}$} & \colhead{$A_V$} & \colhead{Period} & \colhead{Rad} & \colhead{log L$_\nu$} & \colhead{log L$_{X,soft}$} & \colhead{$V_L$} & \colhead{$V_{L,\rm V}$} & \colhead{$V_{L, \rm R}$} \\ 
\colhead{ } & \colhead{ } & \colhead{deg} & \colhead{deg} & \colhead{pc} & \colhead{mag} & \colhead{mag} & \colhead{mag} & \colhead{day} & \colhead{R$_\odot$} & \colhead{erg s$^{-1}$ Hz$_{-1}$} & \colhead{erg s$^{-1}$} & \colhead{ } & \colhead{ } & \colhead{ }
}
\startdata
RSH2 & lam And & 354.39204 & 46.45628 & 25.9 $\pm$ 0.1 & 1.34 & 1.27 & 0.0 & 53.95 & ... & 15.49 $\pm$ 0.07 & 30.72 & 6.0 & 6.0 & ... \\
RSH3 & FF Aqr & 330.15191 & -2.74084 & 211 $\pm$ 2 & 2.21 & 1.09 & 0.27 & 9.19 & 4.05 & 17.26 $\pm$ 0.12 & 32.53 & 3.4 & 3.4 & ... \\
RSH8 & UX Ari & 51.64764 & 28.71461 & 50.5 $\pm$ 0.3 & 2.81 & 1.2 & 0.0 & 6.39 & ... & 16.99 $\pm$ 0.06 & 31.05 & 6.8 & 5.4 & 4.0 \\
RSH9 & VY Ari & 42.18333 & 31.11443 & 41.5 $\pm$ 0.1 & 3.54 & 1.24 & 0.0 & 16.2 & ... & 16.14 $\pm$ 0.05 & 30.94 & 2.1 & 2.1 & ... \\
RSH13 & BM Cam & 76.55055 & 59.02122 & 216 $\pm$ 3 & -0.92 & 1.31 & 0.06 & 82.8 & 23.21 & 16.62 $\pm$ 0.35 & 31.37 & 1.2 & 1.2 & ... \\
RSH17 & BH CVn & 203.69968 & 37.18237 & 46.9 $\pm$ 0.2 & 1.43 & 0.65 & 0.0 & 2.61 & ... & 16.67 $\pm$ 0.03 & 30.64 & 1.9 & 1.9 & ... \\
RSH18 & BM CVn & 200.3841 & 38.88036 & 116.6 $\pm$ 0.4 & 1.72 & 1.45 & 0.0 & 20.51 & 7.41 & 16.68 $\pm$ 0.11 & 31.12 & 4.5 & 4.5 & ... \\
RSH19 & BQ CVn & 194.76546 & 47.15146 & 176 $\pm$ 1 & 1.85 & 1.18 & 0.0 & 18.53 & 5.14 & 17.01 $\pm$ 0.15 & 30.99 & 2.6 & 2.6 & ... \\
RSH31 & V0841 Cen & 218.56591 & -60.40821 & 95.0 $\pm$ 0.2 & 3.47 & 1.38 & 0.02 & 6.0 & 3.16 & 16.76 $\pm$ 0.35 & 31.04 & 2.5 & ... & 2.5 \\
RSH33 & V0988 Cen & 209.39102 & -31.65314 & 126.0 $\pm$ 0.4 & 3.71 & 1.37 & 0.05 & 2.48 & 2.78 & 17.31 $\pm$ 0.04 & 30.87 & 11.1 & 11.1 & 1.8 \\
RSH35 & AY Cet & 19.15075 & -2.50065 & 75 $\pm$ 1 & 0.91 & 1.15 & 0.0 & 76.5 & ... & 16.47 $\pm$ 0.1 & 31.22 & 6.4 & 6.4 & ... \\
RSH40 & V1762 Cyg & 287.10672 & 52.42549 & 72.2 $\pm$ 0.5 & 1.41 & 1.34 & 0.0 & 1.79 & ... & 16.59 $\pm$ 0.05 & 30.7 & 8.5 & 8.5 & ... \\
RSH42 & V1971 Cyg & 305.38789 & 32.31424 & 223 $\pm$ 2 & 0.73 & 1.25 & 0.03 & ... & ... & 16.66 $\pm$ 0.38 & 31.06 & 1.2 & 1.2 & ... \\
RSH45 & DK Dra & 183.92275 & 72.55109 & 149.1 $\pm$ 0.5 & 0.0 & 1.33 & 0.0 & 63.15 & 16.12 & 17.16 $\pm$ 0.05 & 31.44 & 8.3 & 8.3 & ... \\
RSH46 & DR Dra & 263.17063 & 74.22749 & 86 $\pm$ 3 & 1.61 & 1.28 & 0.0 & 26.74 & 7.93 & 16.68 $\pm$ 0.15 & 30.85 & 7.5 & 7.5 & ... \\
RSH52 & V0340 Gem & 110.38672 & 26.15919 & 467 $\pm$ 6 & -1.26 & 1.1 & 0.07 & 35.37 & ... & 17.98 $\pm$ 0.11 & 31.84 & 2.3 & 2.3 & ... \\
RSH53 & sig Gem & 115.82835 & 28.88249 & 36.9 $\pm$ 0.4 & 1.13 & 1.35 & 0.0 & 19.42 & ... & 16.95 $\pm$ 0.05 & 30.97 & 19.0 & 5.0 & 1.0 \\
RSH59 & IL Hya & 141.20403 & -23.82645 & 106.5 $\pm$ 0.4 & 1.72 & 1.2 & 0.0 & 12.87 & 5.75 & 16.39 $\pm$ 0.29 & 31.22 & 2.5 & 2.4 & 2.4 \\
RSH65 & HK Lac & 331.23626 & 47.23471 & 144 $\pm$ 1 & 0.71 & 1.29 & 0.0 & 25.83 & ... & 16.8 $\pm$ 0.11 & 31.36 & 3.7 & 3.7 & ... \\
RSH69 & UZ Lib & 233.09681 & -8.53361 & 214 $\pm$ 1 & 1.8 & 1.19 & 0.32 & 4.77 & 5.62 & 17.12 $\pm$ 0.18 & 31.28 & 2.4 & 2.4 & 1.2 \\
RSH72 & AE Lyn & 120.6488 & 57.27336 & 99.6 $\pm$ 0.3 & 1.34 & 0.82 & 0.0 & 10.1 & ... & 16.57 $\pm$ 0.1 & 31.01 & 4.1 & 4.1 & ... \\
RSH74 & AE Men & 96.41778 & -72.04307 & 293 $\pm$ 2 & 0.54 & 1.28 & 0.06 & 12.03 & ... & 17.24 $\pm$ 0.34 & 31.28 & 1.8 & ... & 1.8 \\
RSH75 & BM Mic & 318.62975 & -30.75766 & 145 $\pm$ 1 & 2.29 & 1.08 & 0.0 & 14.6 & 4.09 & 16.74 $\pm$ 0.14 & 30.92 & 2.6 & 2.6 & 1.0 \\
RSH76 & BN Mic & 318.71931 & -31.18391 & 170 $\pm$ 1 & 1.46 & 1.34 & 0.0 & 54.46 & 7.98 & 16.42 $\pm$ 0.38 & 30.95 & 1.3 & 1.3 & ... \\
RSH78 & V1149 Ori & 85.36177 & 3.77806 & 151 $\pm$ 1 & 0.44 & 1.34 & 0.0 & 52.59 & 12.07 & 16.36 $\pm$ 0.35 & 31.04 & 1.2 & 1.2 & ... \\
RSH81 & V1355 Ori & 90.66822 & -0.86029 & 127.5 $\pm$ 0.4 & 3.35 & 1.21 & 0.0 & 3.86 & 2.67 & 17.1 $\pm$ 0.06 & 31.05 & 8.5 & 8.5 & ... \\
RSH82 & EZ Peg & 349.22193 & 25.71953 & 163.1 $\pm$ 0.4 & 3.17 & 1.16 & 0.0 & 11.68 & 2.75 & 16.76 $\pm$ 0.17 & 31.12 & 2.9 & 2.9 & ... \\
RSH83 & II Peg & 358.76979 & 28.63383 & 39.1 $\pm$ 0.1 & 4.17 & 1.38 & 0.0 & 6.71 & ... & 16.45 $\pm$ 0.02 & 31.08 & 8.8 & 3.6 & 2.5 \\
RSH84 & IM Peg & 343.25934 & 16.84107 & 98 $\pm$ 1 & 0.74 & 1.4 & 0.0 & 24.44 & 11.9 & 17.12 $\pm$ 0.11 & 31.31 & 1.5 & ... & 1.5 \\
RSH86 & KX Peg & 335.63544 & 30.35743 & 118.7 $\pm$ 0.3 & 2.01 & 0.97 & 0.0 & ... & 4.05 & 16.26 $\pm$ 0.26 & 30.99 & 1.7 & 1.7 & ... \\
RSH91 & AR Psc & 20.73691 & 7.42029 & 45.5 $\pm$ 0.1 & 3.79 & 1.14 & 0.0 & 12.34 & ... & 16.21 $\pm$ 0.06 & 30.9 & 2.8 & 2.7 & 1.2 \\
RSH95 & BG Psc & 24.28716 & 20.7 & 152 $\pm$ 1 & 2.6 & 0.9 & 0.0 & ... & 2.94 & 16.95 $\pm$ 0.09 & 30.38 & 5.1 & 5.1 & ... \\
RSH101 & V4138 Sgr & 290.66796 & -20.64337 & 75.4 $\pm$ 0.1 & 1.91 & 1.22 & 0.0 & 62.02 & 5.68 & 16.48 $\pm$ 0.07 & 30.77 & 5.9 & 5.9 & ... \\
RSH106 & PX Ser & 242.40853 & 5.87697 & 192 $\pm$ 1 & 2.51 & 1.26 & 0.14 & 12.25 & 4.12 & 16.48 $\pm$ 0.42 & 30.79 & 1.0 & 1.0 & ... \\
RSH107 & V0711 Tau & 54.1969 & 0.58704 & 29.4 $\pm$ 0.0 & 3.26 & 1.24 & 0.0 & 2.84 & ... & 16.5 $\pm$ 0.01 & 31.12 & 4.3 & 4.3 & 1.6 \\
RSH108 & XX Tri & 30.94602 & 35.59123 & 196 $\pm$ 1 & 1.1 & 1.37 & 0.0 & 24.26 & ... & 16.84 $\pm$ 0.18 & 31.48 & 2.4 & 2.4 & ... \\
RSH110 & DM UMa & 163.93109 & 60.46933 & 186 $\pm$ 1 & 2.57 & 1.3 & 0.0 & 7.48 & ... & 18.07 $\pm$ 0.02 & 31.38 & 18.2 & 18.2 & ... \\
RSH111 & FF UMa & 143.44375 & 62.82775 & 116.0 $\pm$ 0.4 & 2.32 & 1.27 & 0.0 & ... & 4.43 & 17.16 $\pm$ 0.04 & 31.34 & 7.2 & 7.2 & ... \\
RSH112 & FG UMa & 155.44707 & 60.91275 & 203 $\pm$ 1 & 0.74 & 1.27 & 0.01 & 21.7 & 9.19 & 16.69 $\pm$ 0.33 & 31.08 & 1.6 & 1.6 & ... \\
RSH116 & HU Vir & 183.33616 & -9.07968 & 148 $\pm$ 2 & 3.06 & 1.43 & 0.0 & 10.39 & 3.95 & 17.43 $\pm$ 0.07 & 30.89 & 13.8 & 13.8 & ... \\
RSH121 & NSV 2418 & 83.92982 & -29.92828 & 291 $\pm$ 1 & 2.16 & 1.15 & 0.06 & 3.1 & 4.9 & 17.11 $\pm$ 0.24 & 31.11 & 1.8 & 1.8 & ... \\
RSH126 & MX Dra & 183.06188 & 68.88329 & 203 $\pm$ 1 & 3.7 & 1.45 & 0.0 & 6.4 & ... & 17.1 $\pm$ 0.12 & 30.73 & 3.0 & 3.0 & ... \\
RSH141 & V0474 Gem & 116.31829 & 20.38786 & 481 $\pm$ 4 & 1.1 & 1.22 & 0.05 & 12.13 & 7.8 & 17.36 $\pm$ 0.37 & 31.28 & 1.4 & 1.4 & ... \\
RSH144 & V0457 Vir & 198.22327 & 8.39309 & 218 $\pm$ 1 & 1.48 & 1.03 & 0.05 & 9.87 & 5.26 & 17.36 $\pm$ 0.09 & 31.42 & 3.5 & 3.5 & ... \\
RSH148 & V1595 Sco & 252.43576 & -36.40623 & 194 $\pm$ 1 & 1.29 & 1.29 & 0.12 & 31.58 & 7.7 & 16.78 $\pm$ 0.21 & 31.31 & 1.8 & 1.8 & ... \\
RSH160 & FP Psc & 10.95376 & 18.78121 & 440 $\pm$ 4 & 2.38 & 1.24 & 0.18 & 13.74 & 4.25 & 17.68 $\pm$ 0.29 & 31.08 & 2.7 & 2.7 & ... \\
RSH167 & HI Psc & 20.5645 & 20.35842 & 261 $\pm$ 1 & 2.05 & 0.99 & 0.2 & 10.23 & 3.77 & 17.13 $\pm$ 0.19 & 31.17 & 2.3 & 2.3 & ... \\
RSH168 & HR Psc & 24.11589 & 25.14313 & 219 $\pm$ 1 & 3.09 & 1.39 & 0.37 & 3.93 & 3.75 & 17.61 $\pm$ 0.04 & 31.57 & 9.9 & 9.9 & 1.8 \\
RSH178 & AM Hor & 59.90312 & -39.88759 & 137.8 $\pm$ 0.3 & 3.47 & 1.35 & 0.0 & 9.92 & 3.1 & 16.8 $\pm$ 0.14 & 30.9 & 3.4 & 3.4 & ... \\
RSH190 & V1859 Ori & 80.72831 & 8.96793 & 288 $\pm$ 1 & 2.39 & 1.16 & 0.28 & 5.95 & 3.65 & 17.03 $\pm$ 0.29 & 31.51 & 1.6 & 1.6 & ... \\
RSH197 & ASAS J060415+1245.9 & 91.06248 & 12.76417 & 514 $\pm$ 5 & 0.76 & 1.39 & 0.36 & 21.18 & 10.72 & 18.44 $\pm$ 0.06 & 32.44 & 3.2 & 1.5 & 1.6 \\
RSH217 & V0726 Pup & 119.7411 & -35.37146 & 233 $\pm$ 1 & 3.5 & 1.21 & 0.21 & 2.37 & 2.6 & 16.84 $\pm$ 0.29 & 30.93 & 1.4 & 1.4 & ... \\
RSH226 & EE Pyx & 132.5813 & -28.94409 & 757 $\pm$ 9 & 0.35 & 1.27 & 0.2 & 32.64 & ... & 17.91 $\pm$ 0.26 & 32.12 & 1.6 & 1.6 & ... \\
RSH238 & IN Leo & 159.99586 & 13.45601 & 234 $\pm$ 1 & 3.14 & 1.22 & 0.01 & 6.22 & ... & 17.41 $\pm$ 0.15 & 30.95 & 2.9 & 2.9 & 1.0 \\
RSH242 & OS Leo & 173.40398 & 7.85779 & 298 $\pm$ 2 & 3.55 & 1.22 & 0.1 & 5.71 & 2.53 & 17.24 $\pm$ 0.21 & 31.08 & 2.3 & 2.3 & ... \\
RSH248 & PW Com & 188.98908 & 13.49021 & 206 $\pm$ 1 & 3.28 & 1.27 & 0.02 & 4.53 & 3.05 & 16.79 $\pm$ 0.25 & 31.07 & 1.8 & 1.8 & ... \\
RSH250 & V0445 Vir & 196.37144 & 12.82648 & 316 $\pm$ 2 & 2.39 & 1.01 & 0.0 & 3.68 & 3.47 & 17.83 $\pm$ 0.06 & 30.95 & 8.7 & 8.7 & 1.0 \\
RSH258 & NR Lib & 225.52991 & -12.03134 & 206 $\pm$ 1 & 3.93 & 1.41 & 0.32 & 2.68 & 2.67 & 17.83 $\pm$ 0.04 & 30.83 & 19.1 & 19.1 & ... \\
RSH263 & V0454 Ser & 234.14343 & 12.32499 & 286 $\pm$ 1 & 2.92 & 1.32 & 0.08 & 5.41 & 4.06 & 17.13 $\pm$ 0.25 & 30.81 & 2.0 & 2.0 & ... \\
RSH266 & V0474 Ser & 237.17055 & -3.17906 & 482 $\pm$ 4 & 1.5 & 0.99 & 0.66 & 1.92 & 4.75 & 17.98 $\pm$ 0.1 & 31.13 & 4.6 & 4.6 & ... \\
RSH289 & V3084 Oph & 266.60573 & 3.98 & 117.0 $\pm$ 0.2 & 3.82 & 1.37 & 0.0 & 8.45 & 2.43 & 17.2 $\pm$ 0.04 & 30.18 & 10.4 & 10.4 & ... \\
RSH291 & ASAS J180724+1942.4 & 271.85053 & 19.7063 & 241 $\pm$ 1 & 2.1 & 1.5 & 0.01 & 33.31 & 6.17 & 16.95 $\pm$ 0.24 & 31.44 & 2.1 & 2.1 & ... \\
RSH297 & V5651 Sgr & 287.89468 & -34.58591 & 216 $\pm$ 1 & 2.9 & 1.05 & 0.18 & 3.38 & 2.79 & 16.73 $\pm$ 0.29 & 30.84 & 1.3 & 1.3 & ... \\
RSH300 & V5840 Sgr & 292.13331 & -35.13308 & 197 $\pm$ 1 & 2.57 & 0.98 & 0.4 & 3.24 & 3.02 & 17.19 $\pm$ 0.09 & 31.23 & 4.5 & 4.5 & ... \\
RSH310 & CU Cap & 304.76246 & -14.03462 & 208 $\pm$ 1 & 3.4 & 1.38 & 0.2 & 11.78 & 3.16 & 17.18 $\pm$ 0.11 & 31.33 & 3.5 & 3.5 & ... \\
RSH312 & CY Cap & 307.17622 & -9.72141 & 356 $\pm$ 2 & 1.03 & 0.98 & 0.23 & 2.4 & 6.55 & 17.14 $\pm$ 0.41 & 30.93 & 1.4 & 1.4 & ... \\
RSH316 & DO Mic & 311.40336 & -39.83243 & 280 $\pm$ 2 & 2.35 & 1.2 & 0.21 & 12.49 & 4.01 & 16.98 $\pm$ 0.31 & 31.53 & 1.7 & 1.7 & ... \\
RSH325 & CY Ind & 320.18365 & -54.63311 & 200 $\pm$ 1 & 3.81 & 1.34 & 0.03 & 2.4 & 2.72 & 16.89 $\pm$ 0.35 & 30.85 & 1.0 & ... & 1.0 \\
RSH331 & V0378 Aqr & 335.12469 & -1.66617 & 655 $\pm$ 6 & 2.83 & 1.35 & 0.24 & 9.1 & 3.9 & 17.67 $\pm$ 0.44 & ... & 1.2 & 1.2 & ... \\
RSH334 & ASAS J231713+0551.1 & 349.30619 & 5.8522 & 428 $\pm$ 4 & 1.39 & 1.23 & 0.32 & 4.44 & 7.25 & 17.67 $\pm$ 0.15 & 31.24 & 2.8 & 2.8 & ... \\
RSH337 & V0414 Hya & 140.7234 & -13.82247 & 235 $\pm$ 2 & 1.55 & 1.3 & 0.03 & 18.76 & ... & 16.74 $\pm$ 0.44 & 31.57 & 1.1 & 1.1 & ... \\
RSH339 & HK Boo & 217.25442 & 12.12258 & 117.4 $\pm$ 0.5 & 2.84 & 1.16 & 0.0 & 17.98 & 3.31 & 16.98 $\pm$ 0.06 & 30.61 & 8.8 & 8.8 & ... \\
RSH342 & NT Com & 195.94131 & 28.62249 & 213 $\pm$ 1 & 3.81 & 1.15 & 0.0 & 3.2 & 2.04 & 16.76 $\pm$ 0.28 & 30.39 & 2.0 & 2.0 & ... \\
RSH351 & V1214 Her & 251.76435 & 9.76617 & 401 $\pm$ 3 & 3.18 & 1.15 & 0.14 & 4.59 & 2.77 & 17.57 $\pm$ 0.17 & 30.85 & 3.2 & 3.2 & ... \\
RSH356 & V0597 Peg & 353.01927 & 32.46561 & 576 $\pm$ 7 & 2.94 & 1.42 & 0.25 & 7.84 & 4.02 & 17.68 $\pm$ 0.23 & 31.19 & 2.0 & 2.0 & ... \\
RSH358 & V0463 Cam & 103.32545 & 72.76156 & 648 $\pm$ 21 & 2.51 & 1.29 & 0.3 & 5.62 & ... & 18.04 $\pm$ 0.2 & 31.03 & 2.8 & 2.8 & ... \\
RSH362 & V0495 And & 9.2996 & 44.21652 & 584 $\pm$ 6 & 1.27 & 0.96 & 0.14 & 16.71 & 4.82 & 17.45 $\pm$ 0.45 & 31.83 & 1.1 & 1.1 & ... \\
RSH370 & V0486 And & 5.34592 & 33.71028 & 268 $\pm$ 1 & 3.28 & 1.19 & 0.1 & 8.35 & 2.72 & 16.71 $\pm$ 0.43 & 30.82 & 1.0 & 1.0 & ... \\
RSH373 & BY Tri & 35.3888 & 34.07915 & 204 $\pm$ 1 & 2.98 & 0.97 & 0.11 & 3.62 & 2.46 & 16.51 $\pm$ 0.44 & 30.76 & 1.0 & 1.0 & ... \\
RSH374 & CG Ari & 48.88295 & 26.08042 & 393 $\pm$ 3 & 3.52 & 1.43 & 0.64 & 9.32 & 3.17 & 17.28 $\pm$ 0.3 & 31.13 & 1.8 & 1.8 & ... \\
RSH376 & V0941 Per & 60.02428 & 39.69347 & 311 $\pm$ 3 & 3.52 & 1.45 & 0.85 & 5.62 & 3.39 & 17.35 $\pm$ 0.3 & 31.1 & 1.9 & 1.9 & ... \\
RSH378 & V0950 Per & 61.97627 & 35.46368 & 403 $\pm$ 3 & 1.37 & 1.16 & 0.59 & 9.48 & 6.31 & 17.85 $\pm$ 0.08 & 31.49 & 5.7 & 5.7 & ... \\
RSH379 & V1315 Tau & 64.97382 & 30.16487 & 375 $\pm$ 2 & 2.05 & 1.44 & 0.78 & 10.27 & ... & 18.0 $\pm$ 0.06 & 31.3 & 9.9 & 9.9 & ... \\
RSH380 & V1362 Tau & 80.54314 & 24.53571 & 185 $\pm$ 1 & 3.29 & 1.0 & 0.94 & 2.52 & 2.34 & 16.69 $\pm$ 0.22 & 30.94 & 1.8 & 1.8 & ... \\
RSH391 & V0832 Cep & 313.85889 & 61.59115 & 352 $\pm$ 1 & 3.95 & 1.4 & 0.63 & 5.81 & 2.57 & 17.07 $\pm$ 0.38 & 30.96 & 1.2 & 1.2 & ... \\
RSH394 & V0406 Gem & 101.81492 & 14.57754 & 662 $\pm$ 7 & 1.49 & 1.34 & 0.08 & 11.22 & 7.94 & 17.97 $\pm$ 0.19 & 31.52 & 2.1 & 2.1 & ... \\
RSH402 & V0948 Per & 61.06449 & 49.73237 & 274 $\pm$ 1 & 2.97 & 1.09 & 0.53 & 10.03 & 2.87 & 17.35 $\pm$ 0.12 & 30.75 & 4.8 & 4.8 & ... \\
RSH444 & AD LMi & 156.5943 & 37.75369 & 186 $\pm$ 1 & 3.09 & 1.27 & 0.0 & 17.8 & 3.4 & 16.59 $\pm$ 0.29 & 31.19 & 1.5 & 1.5 & ... \\
RSH471 & KZ Psc & 349.18761 & 6.31593 & 278 $\pm$ 2 & 2.85 & 1.27 & 0.37 & 4.19 & 3.84 & 17.43 $\pm$ 0.12 & 30.54 & 4.9 & 4.9 & 1.0 \\
RSH484 & V0545 Dra & 307.41346 & 73.90951 & 216 $\pm$ 1 & 3.67 & 1.37 & 0.23 & 3.57 & 2.84 & 17.28 $\pm$ 0.09 & 31.01 & 5.3 & 5.3 & ... \\
RSH510 & GSC 03642-02459 & 354.58776 & 48.43137 & 418 $\pm$ 2 & 1.9 & 0.98 & 0.25 & 1.86 & 4.27 & 17.25 $\pm$ 0.34 & ... & 1.3 & 1.3 & ... \\
RSH518 & V0974 Cep & 314.37585 & 60.0525 & 330 $\pm$ 1 & 2.67 & 0.95 & 0.46 & 8.24 & 2.82 & 16.95 $\pm$ 0.39 & 30.97 & 1.1 & 1.1 & ... \\
RSH550 & ASAS J050457-0354.8 & 76.23967 & -3.91466 & 131.0 $\pm$ 0.4 & 2.81 & 1.08 & 0.0 & 0.91 & 3.12 & 17.12 $\pm$ 0.07 & 31.14 & 9.0 & 9.0 & ... \\
RSH553 & ASAS J162848-4152.6 & 247.19703 & -41.87756 & 286 $\pm$ 2 & 4.68 & 2.31 & 0.3 & 4.94 & 4.07 & 17.43 $\pm$ 0.26 & 31.09 & 1.6 & ... & 1.6 \\
RSH558 & HAT-187-0002449 & 196.73283 & 36.95227 & 599 $\pm$ 6 & 4.8 & 1.54 & 0.0 & 1.7 & 1.86 & 17.4 $\pm$ 0.44 & ... & 1.0 & 1.0 & ... \\
RSH563 & NSVS 13039638 & 154.67419 & -3.69102 & 727 $\pm$ 17 & 4.22 & 1.48 & 0.09 & 4.12 & ... & 17.7 $\pm$ 0.46 & 31.18 & 1.0 & 1.0 & ... \\
RSH891 & ASASSN-V J022656.00+203659.4 & 36.73336 & 20.61639 & 448 $\pm$ 3 & 3.43 & 1.4 & 0.39 & 11.03 & 3.32 & 17.25 $\pm$ 0.38 & ... & 1.1 & 1.1 & ... \\
RSH968 & EVRJ191908.38+083523.6 & 289.79373 & 8.57884 & 674 $\pm$ 7 & 2.53 & 1.45 & 1.11 & 5.68 & 4.95 & 18.4 $\pm$ 0.07 & ... & 5.5 & 5.5 & ... \\
RSH988 & ZALD J02284341+8235565 & 37.18073 & 82.59902 & 564 $\pm$ 5 & 2.12 & 1.0 & 0.54 & 5.28 & 3.6 & 17.74 $\pm$ 0.21 & 31.0 & 2.7 & 2.7 & ... \\
RSH1007 & ZALD J23315507+6654022 & 352.97943 & 66.9006 & 282 $\pm$ 1 & 2.96 & 1.16 & 0.14 & 34.75 & 3.13 & 16.9 $\pm$ 0.35 & 30.7 & 1.3 & 1.3 & ... \\
RSH1152 & ZTF J002855.40+591924.4 & 7.23085 & 59.32345 & 423 $\pm$ 2 & 3.99 & 1.53 & 0.56 & 5.31 & 2.87 & 17.19 $\pm$ 0.43 & ... & 1.2 & 1.2 & ... \\
RSH1188 & ZTF J004321.77+615450.4 & 10.84076 & 61.91398 & 563 $\pm$ 6 & 4.87 & 2.04 & 0.96 & 3.18 & ... & 17.62 $\pm$ 0.24 & ... & 1.9 & 1.9 & ... \\
RSH2933 & ZTF J181524.21-015718.7 & 273.85089 & -1.95522 & 954 $\pm$ 19 & 2.46 & 1.85 & 3.28 & 5.87 & 6.22 & 18.39 $\pm$ 0.18 & ... & 2.8 & 2.8 & ... \\
RSH3759 & ZTF J191426.01+125823.7 & 288.60839 & 12.97321 & 1178 $\pm$ 46 & 3.33 & 1.6 & 2.36 & 4.1 & 3.64 & 18.02 $\pm$ 0.48 & ... & 1.0 & 1.0 & ... \\
RSH5961 & ZTF J212223.68+330305.9 & 320.59868 & 33.05162 & 933 $\pm$ 11 & 3.59 & 1.4 & 0.27 & 3.81 & ... & 17.9 $\pm$ 0.38 & ... & 1.3 & 1.3 & ... \\
RSH7095 & ZTF J185809.69+022129.5 & 284.54036 & 2.35817 & 422 $\pm$ 3 & 4.07 & 1.52 & 0.33 & 7.39 & 2.7 & 17.38 $\pm$ 0.34 & ... & 1.8 & 1.8 & ... \\
RSH7153 & Gaia DR2 1964619166204043008 & 322.53189 & 39.66103 & 857 $\pm$ 9 & 3.5 & 1.39 & 0.57 & 3.57 & 3.15 & 17.76 $\pm$ 0.41 & ... & 1.1 & 1.1 & ... \\
RSH7181 & ZALD J03461036+6219202 & 56.543 & 62.32224 & 284 $\pm$ 1 & 2.97 & 1.21 & 0.34 & 1.47 & ... & 16.95 $\pm$ 0.34 & 30.99 & 1.3 & 1.3 & ... \\
\enddata
\end{deluxetable*}
\end{longrotatetable}

\end{document}